\documentclass[
  superscriptaddress,
  longbibliography,
  amsmath,
  showkeys,
]{revtex4-2}

\newcommand{\theTitle}{Computational generation of tailored radionuclide libraries for alpha-particle and gamma-ray spectrometry}
\newcommand{\authI}{Jaewoong \surname{Jang}}
\newcommand{\authIEmail}{jang@ric.u-tokyo.ac.jp}
\newcommand{\addrUTokyoISC}{%
  Isotope Science Center,
  University of Tokyo,
  2-11-16 Yayoi,
  Bunkyo,
  Tokyo 113-0032,
  Japan%
}

\usepackage{upgreek}
\usepackage[
  colorlinks=true,
  linkcolor=blue,
  citecolor=blue,
  urlcolor=blue,
]{hyperref}

\usepackage{graphicx}
\graphicspath{{./}}

\newcommand{\eqPunctDot}{.}
\newcommand{\eqPunctComma}{,}
\newcommand{\eqRef}[2][Eq.]{#1~(\ref{#2})}
\newcommand{\figRef}[1]{Fig.~\ref{#1}}
\newcommand{\FigRef}[1]{Figure~\ref{#1}}
\newcommand{\subfigRef}[2]{Fig.~\ref{#1}\textbf{#2}}
\newcommand{\SubfigRef}[2]{Figure~\ref{#1}\textbf{#2}}
\newcommand{\subfigsRef}[3]{Figs.~\ref{#1}\textbf{#2} and \ref{#1}\textbf{#3}}
\newcommand{\secRef}[2][.]{Sec#1~\ref{#2}}
\newcommand{\CRediTAuth}[1]{\textbf{#1}}

\begin{document}

\title{\theTitle}
\author{\authI}
\email[]{\authIEmail}
\affiliation{\addrUTokyoISC}
\date{\today}

\begin{abstract}
  Radionuclide identification is a radioanalytical method employed in various scientific disciplines that utilize alpha-particle or gamma-ray spectrometric assays, ranging from astrophysics to nuclear medicine. Radionuclide libraries in conventional radionuclide identification systems are crafted in a manual fashion, accompanying labor-intensive and error-prone user tasks and hindering library customization. This research presents a computational algorithm and the architecture of   its dedicated software that can automatically generate tailored radionuclide libraries. Progenitor-progeny recurrence relations were modeled to enable recursive computation of radionuclide subsets. This theoretical concept was incorporated into open-source software called \texttt{RecurLib} and validated against four actinide decay series and twelve radioactive substances, including a uranium-glazed legacy Fiestaware, natural uranium and thorium sources, a \textsuperscript{226}Ra sample, and the medical radionuclides \textsuperscript{225}Ac, \textsuperscript{177}Lu, and \textsuperscript{99m}Tc. The developed algorithm yielded radionuclide libraries for all the tested specimens within minutes, demonstrating its efficiency and applicability across diverse scenarios. The proposed approach introduces a framework for computerized radionuclide library generation, thereby trivializing library-driven radionuclide identification and facilitating the spectral recognition of unregistered radionuclides in radiation spectrometry.
\end{abstract}


\maketitle

\section{Introduction}

Identifying the radionuclidic compositions of radioactive substances holds significant importance across various fields utilizing radiation spectrometry, including astronomy \cite{RN732,RN727,RN736,RN745}, particle physics \cite{RN730,RN729}, nuclear physics \cite{RN737,RN738,RN744,RN739,RN749,RN746}, geology \cite{RN799,RN807,RN735,RN726}, environmental monitoring \cite{RN775,RN770,RN771,RN769}, materials science \cite{RN740,RN743,RN728}, homeland security \cite{RN753,RN747}, and nuclear medicine \cite{RN768,RN690,RN685,RN510,RN742,RN665,RN733,RN734}. Radionuclide identification is performed through spectral analysis of alpha particles or gamma radiation using their discrete energies. Once identified, radionuclides intended for use can be further assessed according to the applications of interest, and those posing potential threats can be screened for nuclear trafficking at border checkpoints. Alpha-particle and gamma-ray spectrometry are commonly employed to meet these needs.

Substantial efforts have been dedicated to developing radionuclide identification methodologies. The majority of these approaches depend on either radionuclide libraries \cite{RN367,RN776} or pattern recognition \cite{RN780,RN779,RN781,RN782}, with the former being increasingly used in daily spectral assays and the latter being broadly investigated with the growing use of artificial intelligence (AI). Although AI-assisted predictive modeling shows promise for developing rapid radionuclide identification systems, extensive training is still required to achieve adequate generalization capabilities. At present, the conventional library-driven methods remain the standard of practice in laboratory settings.

Accurate radionuclide identification relies heavily on the availability and comprehensiveness of radionuclide libraries. A radionuclide library is a compilation of radionuclides and their associated nuclear data, encompassing the type, energy, and emission probability of decay radiation, and the radionuclidic half-life. In general, constructing a radionuclide library involves a series of manual tasks: (i) enumerating the candidate radionuclides that are suspected to be present in the spectral sample under consideration, (ii) acquiring their nuclear data from an authoritative nuclear database, and (iii) linking the fetched data to the corresponding radionuclides. In commercial spectral analysis software, this process is accomplished by pulling radionuclides from a precompiled master nuclear library \cite{RN776}.

Despite the widespread utilization and significance of radionuclide libraries in spectral analysis, insufficient consideration has been paid to their automation. Most studies addressing library-based radionuclide identification seek to develop algorithms for peak locating \cite{RN792,RN788,RN787} or multiplet deconvolution \cite{RN786,RN800,RN794} to improve true peak recognition rates, leaving the construction of the required radionuclide libraries to end users. Without a means of automation, such library generation involves laborious and error-susceptible tasks; not only does this lead to difficulties in building libraries for newly encountered radionuclides, but it can also result in inaccurately registered nuclear data by human error. Implementing computerization in library generation can alleviate the workload on users, improve the library data reliability, and streamline the process of radionuclide identification.

The aim of this study was to introduce automation into the process of radionuclide library creation, thereby improving the efficiency and applicability of library-guided radionuclide identification approaches while minimizing human intervention. This was achieved through the development of a recursive algorithm and its dedicated software, which were validated using radiation spectra involved in geological sciences and nuclear medicine. This paper provides comprehensive accounts of the computational algorithm and its software architecture, and explores the task reduction in radionuclide identification facilitated by computerized libraries. Special emphasis is placed on the radiological characterization of naturally occurring radioactive materials (NORM), whose spectral complexity and interdisciplinary significance effectively demonstrate the broad applicability of the algorithm.

\section{Theoretical conceptualization}

\subsection{Computational radionuclide subsets}

Let us consider a case where a parent radionuclide $d_{0,0}$ decays into its daughter radionuclide $d_{0,1}$, which produces its own daughter radionuclide $d_{0,2}$. Assuming that such a chain of nuclear transmutation continues until a stable nuclide $d_{0,n}$ is reached by the decay of its parent radionuclide $d_{0,n-1}$, the sequence of all radionuclides involved can be expressed as:
\begin{equation}
  \label{eq:manySetsUnion}
  D_{0} = \left\{ d_{0,0},\, d_{0,1},\, d_{0,2},\, \cdots,\, d_{0,n-1} \right\}
  \eqPunctDot
\end{equation}
The first element in such a set, or $d_{0,0}$ in this particular instance, is hereafter referred to as a progenitor radionuclide.

\begin{figure*}
  \centering
  \includegraphics[width=\linewidth]{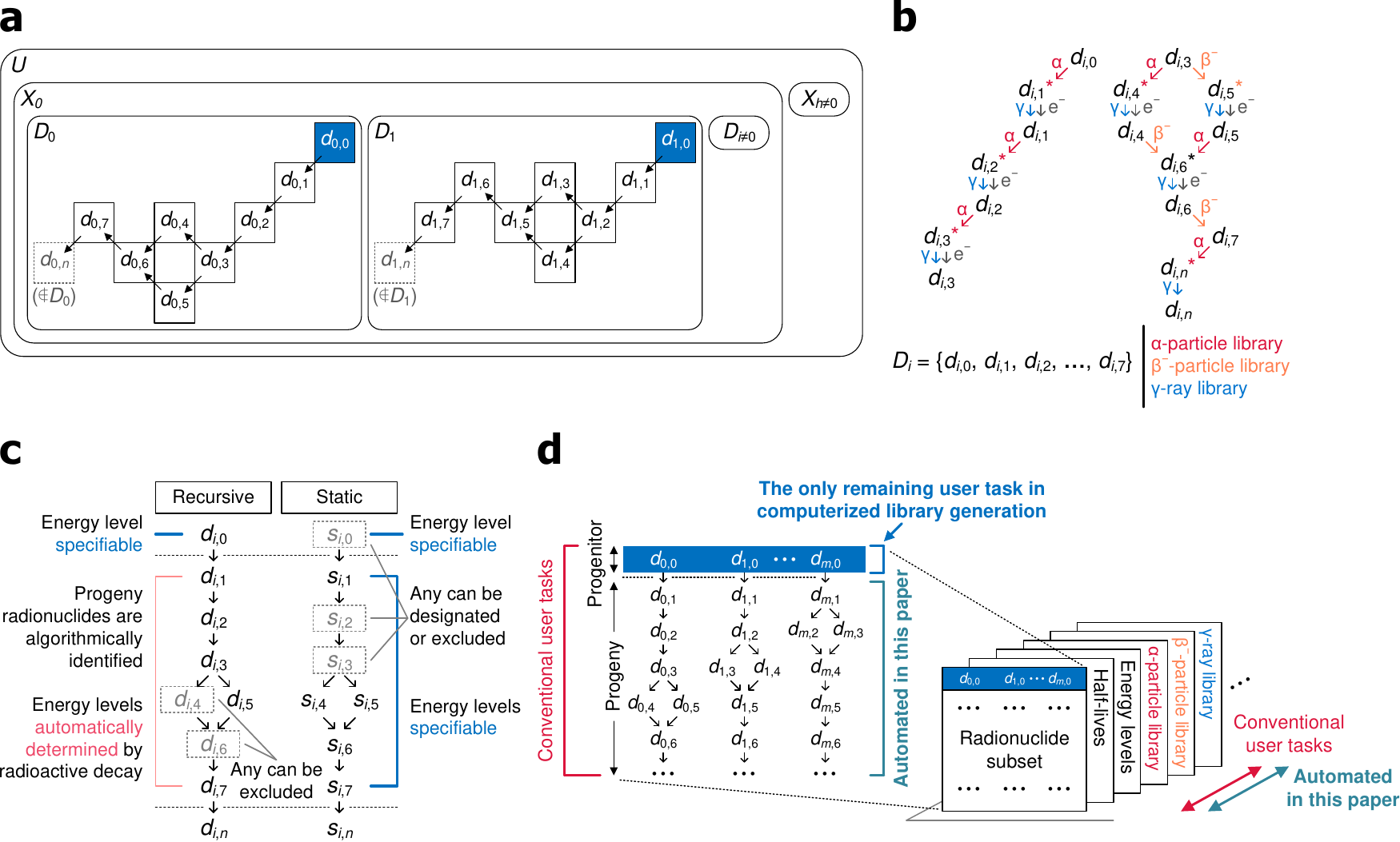}
  \caption{\label{fig:concept}%
    Conceptual illustrations of computational radionuclide library generation. \textbf{a} Logical relations between member radionuclides ($d_{0,0}$, $d_{0,1}$, $\cdots$, $d_{0,7}$ and $d_{1,0}$, $d_{1,1}$, $\cdots$, $d_{1,7}$), decay chains ($D_{0}$, $D_{1}$, and $D_{i \neq 0,1}$), radionuclide subsets ($X_{0}$ and $X_{h \neq 0}$), and a universal set ($U$). \textbf{b} Multiple radiation types can be involved in a single decay chain. \textbf{c} Recursive and static decay chains are schematically compared. \textbf{d} Most manual tasks in the conventional library construction approach are computerized by the developed algorithm.%
  }
\end{figure*}

Radiation spectrometry samples, even those that have been chemically purified, can contain more than one progenitor radionuclides; examples include the therapeutic radionuclide \textsuperscript{225}Ac where measurable quantities of \textsuperscript{227}Ac \cite{RN685} or \textsuperscript{226}Ac \cite{RN665} can coexist depending on the production method, and NORM specimens in which \textsuperscript{238}U, \textsuperscript{235}U, \textsuperscript{232}Th, \textsuperscript{40}K, and their progeny can reside together \cite{RN770,RN771,RN769}. \eqRef{eq:manySetsUnion} can therefore be generalized as
\begin{align}
  \label{eq:manySetsUnionGeneralized}
  X_{h} &= \bigcup_{i=0}^{m} D_{i} \\ \nonumber
  &= \bigcup_{i=0}^{m} \bigcup_{j=0}^{n-1} \left\{ d_{i,j} \right\}
  \eqPunctComma
\end{align}
of which the logical relations are summarized in \subfigRef{fig:concept}{a} using arbitrary decay chains $D_{0}$ and $D_{1}$.

It is the radionuclide subset $X_{h}$ against which, with the corresponding nuclear data available, given radiation spectra can be compared for radionuclide identification. The universal set $U$, in this context, is a set containing the names of all known radionuclides, from which $X_{h}$ can be drawn by selectively retrieving singletons of $\{ d_{i,j} \}$.

However, constructing $X_{h}$ using \eqRef{eq:manySetsUnionGeneralized} in its present form requires manual identification of every progeny nuclide $d_{i,j \neq 0}$ for each progenitor $d_{i,0}$, which is time-consuming and error-prone. To facilitate its computation, I rewrite \eqRef{eq:manySetsUnionGeneralized} in matrix notation:
\begin{align*}
  \mathbf{X_{h}} &= \begin{bmatrix}
    D_{0} \\
    D_{1} \\
    \vdots \\
    D_{m}
  \end{bmatrix} \\
  &= \begin{bmatrix}
    d_{0,0} & d_{0,1} & d_{0,2} & \cdots & d_{0,n-1} \\
    d_{1,0} & d_{1,1} & d_{1,2} & \cdots & d_{1,n-1} \\
    \vdots & \vdots & \vdots & \ddots & \vdots \\
    d_{m,0} & d_{m,1} & d_{m,2} & \cdots & d_{m,n-1} \\
  \end{bmatrix}
  \eqPunctDot
\end{align*}
A computationally efficient approach would be to separate a set of progenitor radionuclides from their respective progeny and subsequently compute the latter:
\begin{align}
  \label{eq:manySetsUnionGeneralizedMatrixComputational}
  \mathbf{X_{h}} &= \begin{bmatrix}
    d_{0,0} & 0 & 0 & \cdots & 0 \\
    d_{1,0} & 0 & 0 & \cdots & 0 \\
    \vdots & \vdots & \vdots & \ddots & \vdots \\
    d_{m,0} & 0 & 0 & \cdots & 0 \\
  \end{bmatrix}
  + \begin{bmatrix}
    0 & d_{0,1} & d_{0,2} & \cdots & d_{0,n-1} \\
    0 & d_{1,1} & d_{1,2} & \cdots & d_{1,n-1} \\
    \vdots & \vdots & \vdots & \ddots & \vdots \\
    0 & d_{m,1} & d_{m,2} & \cdots & d_{m,n-1} \\
  \end{bmatrix} \\ \nonumber
  &= \mathbf{R_{h}} + \mathbf{Y_{h}}
  \eqPunctComma
\end{align}
where $\mathbf{R_{h}}$ and $\mathbf{Y_{h}}$ represent progenitor and progeny matrices, respectively.

Evidently, an algorithm to obtain $\mathbf{Y_{h}}$ should be dependent on $\mathbf{R_{h}}$. I therefore define a function relating a progenitor $d_{i,0}$ to its progeny set $\left\{ d_{i,1}, d_{i,2}, \cdots, d_{i,n - 1} \right\}$, particularly a function $f(j)$ collecting all possible decay products of $j$ recursively until immediately before the stable nuclide $n$ is observed:
\begin{widetext}
\begin{align}
  \label{eq:resursiveFunc}
  f(j) &= g(j) \cup f(j + 1) \\ \nonumber
  &= g(j) \cup g(j + 1) \cup f(j + 2) \\ \nonumber
  &= g(j) \cup g(j + 1) \cup g(j + 2) \cup \cdots \cup g(n - 2) \cup f(n - 1)
  \eqPunctComma
\end{align}
\end{widetext}
where $g(j)$ denotes a set of one or more daughter nuclides into which $j$ transmutes by radioactive decay, which can be expressed as $\{j + 1, j+ 2, \cdots\}$. Meanwhile, the base case of \eqRef{eq:resursiveFunc} will be established when $j = n - 1$ culminates in the stable nuclide $n$, resulting in an empty daughter set $g(j) = \{\}$. The base and recursive cases can then be summarized as
\begin{equation}
  \label{eq:recurlibFuncGeneral}
  f(j) = \begin{cases}
    g(j), & \text{if } g(j) = \{\}, \\
    g(j) \cup f(j + 1), & \text{if } g(j) \neq \{\}.
  \end{cases}
\end{equation}

More than one type of decay radiation can be involved in a single radionuclide subset (\subfigRef{fig:concept}{b}). For example, a decay chain starting with \textsuperscript{225}Ac as the progenitor and ending with \textsuperscript{205}Tl as the stable nuclide involves all the common types of decay radiation: alpha and beta particles, gamma rays, and conversion and Auger electrons, each requiring a dedicated library. Such a radionuclide subset should therefore be processed for all types of the involved decay radiation. Note that by convention, radiation of a nuclide $d_{i,j}$ emitted from its parent-induced excitation state $d_{i,j}$\textsuperscript{*} is ascribed to the parent $d_{i,j-1}$, and so is in nuclear data.

A radionuclide subset should also be able to include, in addition to decay chains computed by \eqRef{eq:recurlibFuncGeneral}, individual radionuclides with their descendants excluded (\subfigRef{fig:concept}{c}). Inclusion of such static radionuclides in a radionuclide subset can be useful when their decay products should be ignored; parent radionuclides whose daughters have been eluted and yet been regrown are the primary subject of such a static group. If static radionuclides are artificially produced nuclear isomers \cite{RN738,RN747,RN749,RN746} or excited states \cite{RN753}, the energy levels need to be explicitly specified. This is in contrast to recursively computed progeny radionuclides, whose energy levels are automatically determined based on the corresponding nuclear data (\secRef{sec:nrgLevFeasVal}). Additionally, library customization may require exclusion of certain radionuclides from either recursive or static family. A complete radionuclide subset enriched by a static radionuclide matrix $\mathbf{S_{h}}$ and subtracted by an exclusion radionuclide matrix $\mathbf{E_{h}}$ can then be defined as
\begin{equation}
  \label{eq:rnSubsetMatrix}
  \mathbf{X_{h}} = \mathbf{R_{h}} + \mathbf{Y_{h}} + \mathbf{S_{h}} - \mathbf{E_{h}}
\end{equation}
using \eqRef{eq:manySetsUnionGeneralizedMatrixComputational} or, equivalently,
\begin{equation*}
  X_{h} = \left( R_{h} \cup Y_{h} \cup S_{h} \right) \setminus E_{h}
  \eqPunctDot
\end{equation*}

\subfigRef{fig:concept}{d} illustrates the task reduction offered by computational radionuclide subsets. By incorporating \eqRef{eq:rnSubsetMatrix} into a software environment, decay chains of radionuclides can be self-assembled into nested data structures, to which nuclear data can be coupled in a programmable fashion. This computability can in turn introduce automation into the manually practiced conventional library generation, leaving progenitor designation the only user task.

\subsection{Nuclear data retrieval and coupling}

Fully automated library construction necessitates a programmed method for nuclear data collection and coupling. Built on top of the radionuclide subsetting algorithm, this system completes the concept of computational radionuclide library generation. The canonical approach in spectral analysis software is to provide users with a precompiled master nuclear library \cite{RN776} from which radionuclide libraries can be withdrawn (\subfigRef{fig:concept}{d}); data obsolescence and inadequacy in such vendor-provided master libraries are not uncommon, requiring manual updates by experienced radiation spectrometrists. Managing master libraries is a complex undertaking, and its editability increases data corruption vulnerability. Instead of fetching nuclear data from such a user-editable database, I propose using a computerized nuclear data retrieval system that accesses an online nuclear database through its web application programming interface (API). This feature can eliminate the need for master library management by end users, and ensures the use of latest evaluated nuclear data.

\section{Software implementation}

\subsection{Architecture}

Open-source software called \texttt{RecurLib} \cite{RN809,RN810} was developed to translate the theoretical framework of algorithmic radionuclide library generation into a practical application. \texttt{RecurLib} is written in Python 3 and works both as a standalone program and as a class that can be integrated into other Python environments.

\begin{figure*}
  \centering
  \includegraphics[width=.55\linewidth]{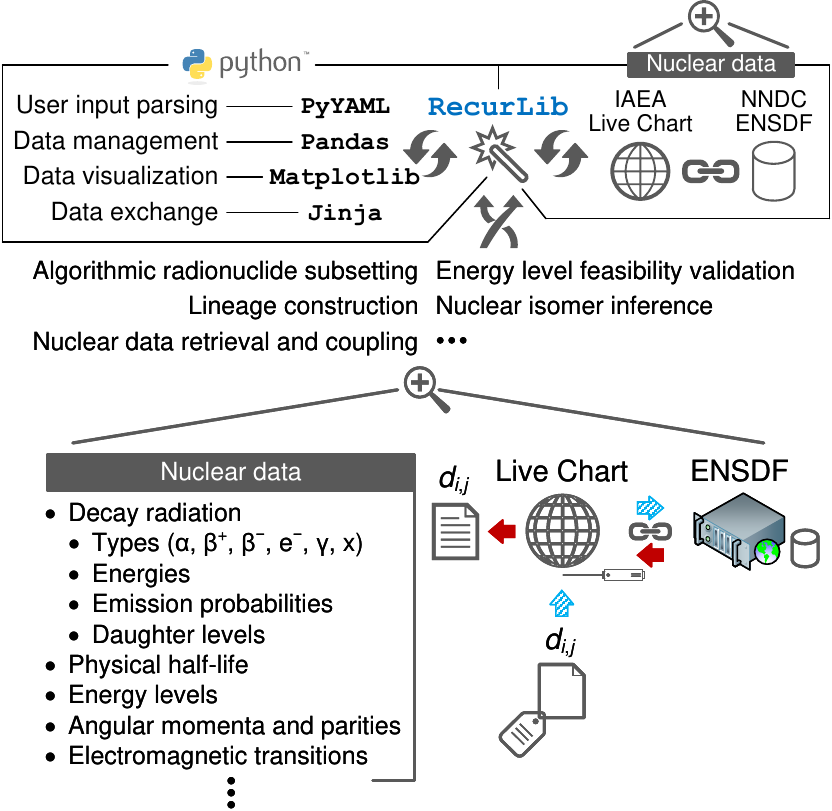}
  \caption{\label{fig:architectureOvv}%
    Third-party dependencies of the developed software \texttt{RecurLib}. Nuclear datasets are retrieved and coupled to \texttt{RecurLib}-constructed radionuclide subsets using the Evaluated Nuclear Structure Data File (ENSDF) of the National Nuclear Data Center (NNDC) via the Live Chart of Nuclides of the International Atomic Energy Agency (IAEA).%
  }
\end{figure*}

The architecture of \texttt{RecurLib} and its dependencies on third-party software are outlined in \figRef{fig:architectureOvv}. Within the Python ecosystem, \texttt{RecurLib} relies on \texttt{PyYAML} \cite{RN764} for user input parsing, \texttt{Pandas} \cite{RN766} for data restructuring and management, \texttt{Matplotlib} \cite{RN767} for data visualization, and \texttt{Jinja} \cite{RN765} for cross-platform data exchange. All these third-party Python libraries are distributed under permissive free software licenses granting the use and redistribution rights free of charge.

The core database of \texttt{RecurLib} is the Evaluated Nuclear Structure Data File (ENSDF) maintained by the National Nuclear Data Center of the Brookhaven National Laboratory \cite{RN761,RN777}; the ENSDF is one of the most authoritative and comprehensive nuclear databases. In practice, access to the ENSDF is mediated by the Live Chart of Nuclides, an interactive web interface developed by the Nuclear Data Section of the International Atomic Energy Agency \cite{RN760}. The Live Chart of Nuclides exposes a web API in addition to its mainstream browser interface, enabling programmatic access to the ENSDF from within external software. This requires one-time Internet connection for newly encountered nuclear datasets (\secRef{sec:rnSubsetConst}).

\subsection{Radionuclide subset construction}
\label{sec:rnSubsetConst}

The computational construction of a radionuclide subset is accomplished by a generative recursive subroutine based on the parent-daughter recurrence relation defined in \eqRef{eq:recurlibFuncGeneral}. In this function, an input parent radionuclide is concatenated with six types of decay radiation ranging from the alpha particle to characteristic X ray, creating radionuclide-radiation pairs. Each pair represents a unit for querying a decay radiation dataset to the Live Chart of Nuclides.

\begin{figure*}
  \centering
  \includegraphics[width=\linewidth]{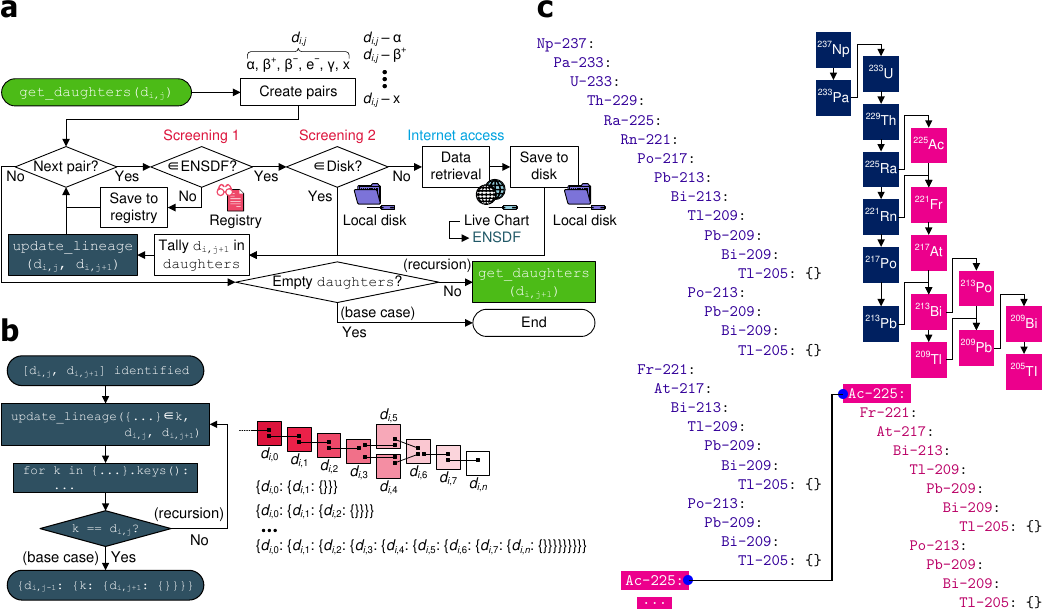}
  \caption{\label{fig:rnSubsettingAndLineageConst}%
    Flowcharts for \textbf{a} radionuclide subsetting and \textbf{b} lineage construction. \textbf{c} A computationally generated \textsuperscript{237}Np lineage.%
  }
\end{figure*}

For computational efficiency, individual queries are executed only if they pass two screening tests (\subfigRef{fig:rnSubsettingAndLineageConst}{a}). The initial screening involves scanning a registry file to determine whether the nuclear dataset in question is included in the ENSDF. Absence rejects access to the Live Chart of Nuclides, prompting the function to log the name of the dataset in the registry file and conclude the query. Passing this first test initiates a secondary inspection, which searches for the corresponding decay data file on the disk; an existing file then eliminates the need for online data retrieval. If unavailable, the file is fetched from the Live Chart of Nuclides and saved to the disk for future reuse. The daughter information is then extracted from the decay data file and tallied in a dedicated sequence, and passed to a lineage formation function (\secRef{sec:lineageConst}).

After processing the last pair, the program proceeds to a block of code stating its recursive and base cases. If the collected daughter set is nonempty, a recursive function is invoked for each daughter element, creating a call stack. A stable nuclide constitutes the base case, which is manifested by an empty set of daughters. This recursive identification of progeny is performed on every designated progenitor. The resulting decay chains of the progenitors are then merged with a group of static radionuclides and subtracted from a set of exclusion radionuclides, if any. This produces a complete radionuclide subset, which is subsequently granted either alpha-particle, beta-particle, or gamma-ray dataset to generate a radionuclide library.

\subsection{Lineage construction}
\label{sec:lineageConst}

\texttt{RecurLib} computes the decay chain of a progenitor as well as individual parent-daughter relations, which is visualized in a lineage diagram. A radionuclide subset generated by \eqRef{eq:recurlibFuncGeneral} is essentially a serial sequence, and this flattened data structure obscures the presence of potential decay branches. In contrast, a lineage tree can illustrate all involved structural relations within the same framework, in which both the branched decay of a parent and the production of a daughter from multiple daughters are explicitly presented. \subfigRef{fig:rnSubsettingAndLineageConst}{b} shows the lineage formation algorithm of \texttt{RecurLib} implemented during radionuclide subset construction. Each time the relation of a parent $d_{i,j}$ and its daughter $d_{i,j+1}$ is established, they are passed to a structural recursive function that associates the pair to a nested data structure using a canonical tree traversal technique. Such lineage information is computed for every recursive progenitor specified in user input, and saved individually as an indented tree to a plain text file (\subfigRef{fig:rnSubsettingAndLineageConst}{c}).

\subsection{Energy level feasibility validation}
\label{sec:nrgLevFeasVal}

\begin{figure*}
  \centering
  \includegraphics[width=.8\linewidth]{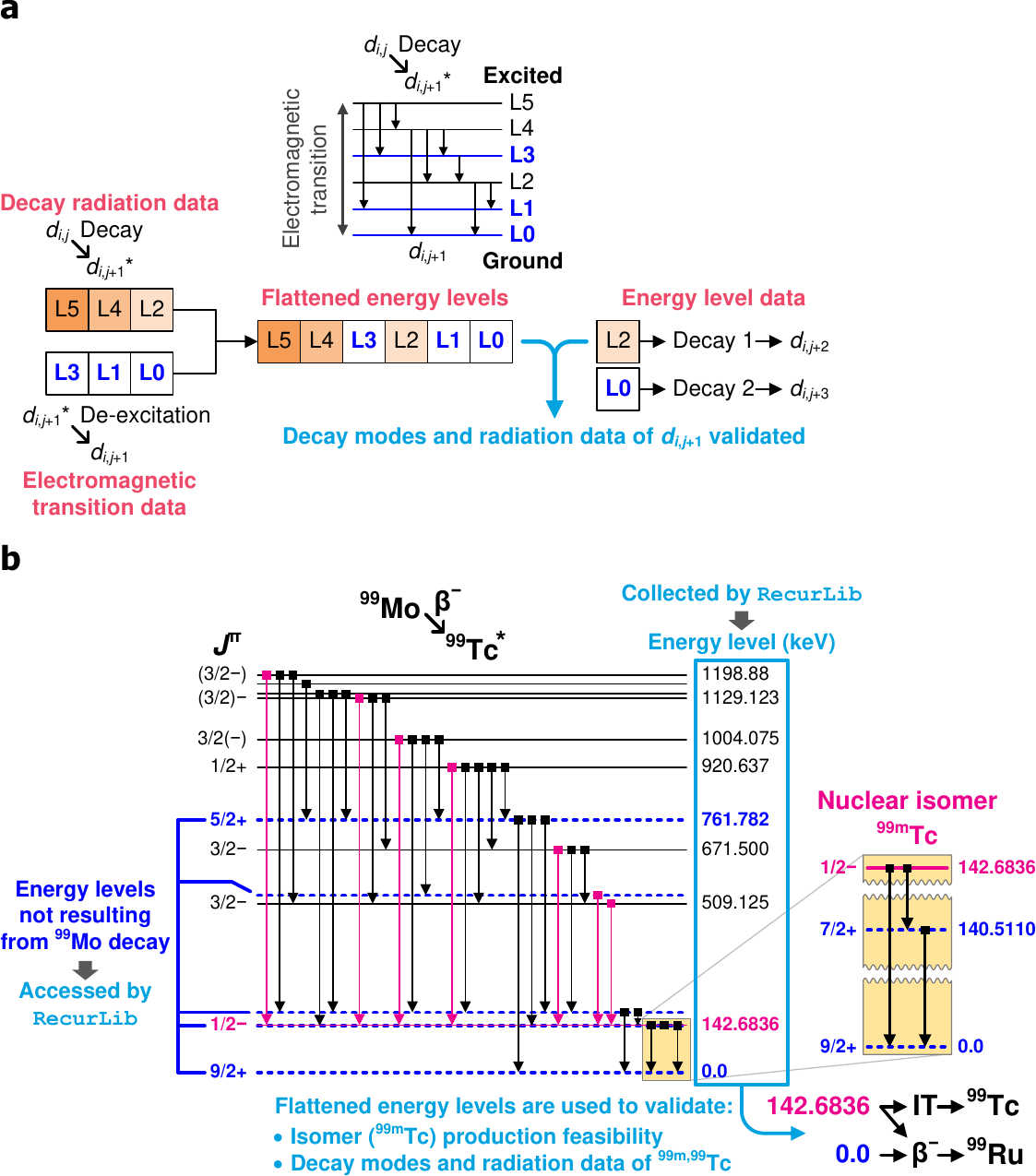}
  \caption{\label{fig:nrgLevVal}%
    Graphical representations of energy level feasibility validation. Schema plots of arbitrary parent-daughter pairs of (\textbf{a}) $d_{i,j}$/$d_{i,j+1}$ and (\textbf{b}) \textsuperscript{99}Mo/\textsuperscript{99}Tc. The asterisk flag, $J^{\pi}$, and IT stand for an excited state, angular momentum-parity pair, and isomeric transition, respectively.%
  }
\end{figure*}

The assurance of data reliability for a radionuclide necessitates the comprehensive consideration of all its permissible energy levels, including both those inherited from parent radionuclides and those subsequently achieved through the cascades of electromagnetic transitions. This concept is schematically illustrated in \subfigRef{fig:nrgLevVal}{a}. The energy levels L5, L4, and L2 belonging to the radionuclide $d_{i,j+1}$ are the immediate results of the decay of its parent $d_{i,j}$. Upon the stabilization of $d_{i,j+1}$, the energy levels L3, L1, and L0 are successively reached by the electromagnetic transitions from L5, L4, or L2; in other words, the levels L3, L1, and L0 are achieved only within $d_{i,j+1}$ after $d_{i,j}$ decays into $d_{i,j+1}$. The set of all energy levels from L5 to L0 is then compared against an energy level dataset presenting at which energy levels $d_{i,j+1}$ undergoes radioactive decay, confirming the allowed decay modes. Such a series of validation process, which I refer to as energy level feasibility validation, plays a vital role in ensuring the integrity of a radionuclide library. For example, if the transition-visited energy levels L3, L1, and L0 are unaccounted for, the corresponding radiation data will be omitted in the radionuclide library and the decay mode at the ground state L0 will be determined unfeasible, leading to a broken radionuclide library.

The energy level feasibility validation in \texttt{RecurLib} is carried out by simulating electromagnetic transitions with the statistical uncertainties considered. Given an arbitrary progeny radionuclide $d_{i,j+1}$, the energy levels originating from its parent $d_{i,j}$ are collected from the nuclear dataset of $d_{i,j}$. \texttt{RecurLib} then traverses the electromagnetic transition dataset of $d_{i,j+1}$ using each energy level inherited from $d_{i,j}$ as input, and keeps track of the visited energy levels. The navigation is performed by a function where an end energy level simulated from its start counterpart is fed as input recursively until the lowest energy is observed; every instance of the call stack tallies both the start and end levels excluding duplicates. This cascade simulation yields the transition-visited energy levels of $d_{i,j+1}$ which, together with the energy levels passed down from $d_{i,j}$, are stored into a sequence. Every radionuclide, either progenitor or progeny, in a radionuclide subset is assigned such a list of flattened energy levels. Finally, individual radionuclides populated with flattened energies undergo energy level feasibility assessment, whereby dropping unviable energy levels and decay modes and generating an integral radionuclide library.

It should be noted that, as progenitors are placed at the top of respective decay chains, they do not possess parent-inherited energy levels provided by \texttt{RecurLib}. Instead, their energy levels are designated at user input; one or more energy levels are specifiable, and omission is cast to the ground state by default. A single nuclide can possess multiple isomeric states and, therefore, energy level specification is required when a nuclear isomer is to be designated as a progenitor.

\subsection{Nuclear isomer inference}
\label{sec:IsomerInference}

The validation of energy level feasibility is also utilized to identify nuclear isomer production. In its current version \cite{RN760}, the Live Chart of Nuclides merges nuclear isomers into their corresponding nuclides, which can only be accessed by their characteristic energy levels. Therefore, assessing the isomer production feasibility of a progeny nuclide necessitates compiling all of its plausible energy levels and consulting the energy level data, which is exactly the same set of information required in the energy level feasibility validation. For this reason, nuclear isomer inference is performed simultaneously when a radionuclide is investigated for its permissible energy levels and decay modes.

\subfigRef{fig:nrgLevVal}{b} depicts a schema plot of \textsuperscript{99}Tc, illustrating the algorithm of nuclear isomer identification and the influence of transition dataset traversal on library integrity. The nuclear isomer of \textsuperscript{99}Tc, or \textsuperscript{99m}Tc, is directly attained at 142.6836 keV by the decay of its parent radionuclide \textsuperscript{99}Mo. Comparison against the energy level dataset of \textsuperscript{99}Tc reveals that the isomeric energy level induces both isomeric transition and beta minus decay. The presence of \textsuperscript{99m}Tc and its two decay modes will therefore be determined to be feasible during energy level validation. On the other hand, five energy levels of \textsuperscript{99}Tc are achievable only through its transition cascades: 761.782 keV, 534.44 keV, 181.094 keV, 140.511 keV, and 0 keV (shown as dashed lines), the last two of which are of practical importance. The state of 140.511 keV with the angular momentum-parity pair of 7/2+ corresponds to the gamma-ray energy used for \textsuperscript{99m}Tc radiopharmaceutical scans \cite{RN234,RN515}; without the cascade traversal simulation, this important gamma-ray energy will be lost in the \textsuperscript{99m,99}Tc gamma-ray library. The 0-keV energy level is also indispensable, as this represents the ground state of \textsuperscript{99}Tc that poses environmental hazards \cite{RN602,RN600}.

\subsection{Integration into external software}

\begin{figure*}
  \centering
  \includegraphics[width=.7\linewidth]{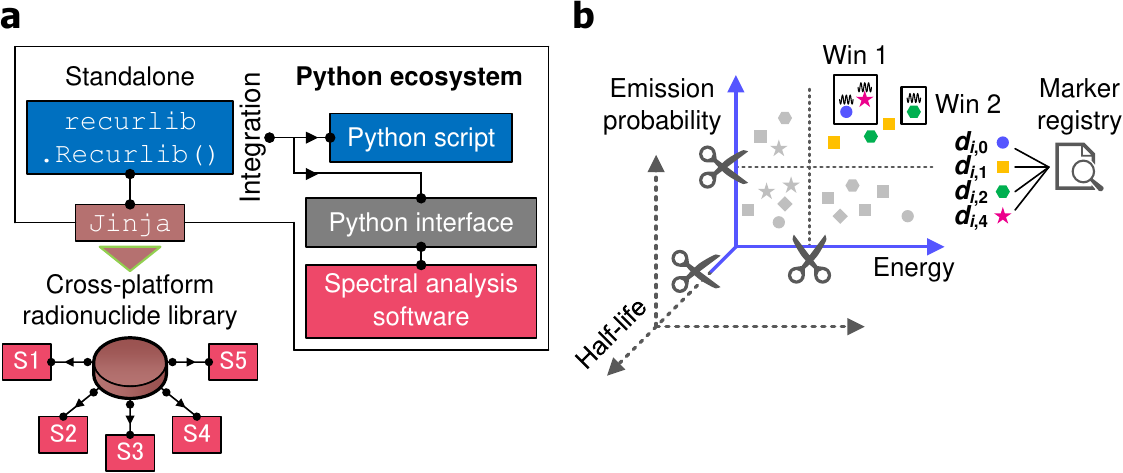}
  \caption{\label{fig:architectureSplit}%
    Architectural features of \texttt{RecurLib}. \textbf{a} Python integration and cross-platform data exchange with arbitrary external software (S1--5). \textbf{b} Library trimming and plot customization using a marker registry and windowing.%
  }
\end{figure*}

In addition to its standalone mode, \texttt{RecurLib} supports two ways of integrating into external software platforms: a modular interface and cross-platform data exchange (\subfigRef{fig:architectureSplit}{a}). When imported into other programs, the \texttt{RecurLib} module provides a radionuclide library class, whose object internally processes the user input and populates its library attribute with a \texttt{Pandas DataFrame} instance. The object can then be further manipulated within the caller. This integration approach can be leveraged to the maximum extent possible within the Python ecosystem, offering a high degree of customization. The second means of software integration is to translate \texttt{RecurLib}-generated radionuclide libraries into cross-platform formats using \texttt{Jinja}, a Python-coded templating engine. By preparing \texttt{Jinja} templates with respect to the data exchange format of the external software of interest, users can obtain exportable radionuclide library files.

For demonstration purposes, I created \texttt{Jinja} templates for use with \texttt{Gamma Explorer} (Mirion Technologies (Canberra) KK, Japan), a spectral analysis software platform widely used in Japan, and generated gamma-ray libraries of four actinide decay series using the templates. These \texttt{RecurLib}-generated cross-platform libraries were all effectively imported into \texttt{Gamma Explorer} (version 1.74) without errors (Supplementary Figures 1--4).

\subsection{Library pruning and marker registration}

\texttt{RecurLib}-generated radionuclide libraries can be pruned by the upper and lower bounds of three nuclear parameters: energy, emission probability, and physical half-life (\subfigRef{fig:architectureSplit}{b}). Energy truncation can serve as a valuable tool in defining valid energy ranges, especially for gamma-ray spectrometric analysis where the deconvolution of complex spectral interferences from characteristic X rays and low-energy gamma rays becomes challenging \cite{RN775,RN367,RN774}. Another practical cutoff criterion is the emission probability; by placing a certain threshold, say 0.001\%, the chance of false positive identification of radionuclides can be substantially decreased \cite{RN367,RN776}. The user can also set a limit on the half-life domain to minimize misidentification of radionuclides. For example, a radionuclide whose half-life is excessively short for detection in plain radiation spectrometry can be excluded from a library, obviating potential confusion with the legitimate radionuclide.

Concepts of marker registration and plot windowing were devised for the coherent representations of radionuclides across different library plots (\subfigRef{fig:architectureSplit}{b}). Users can tailor the marker properties of individual radionuclides using a marker registry file, which was designed to work with the plotting engine \texttt{Matplotlib}. Radionuclides enrolled in a marker registry appear consistent regardless of which library is being drawn, even between alpha-particle and gamma-ray libraries. The windowing feature enables selective annotations of radionuclides within the same canvas along the axes of energy and emission probability. Virtually any number of windows can be created by user input, providing a means of reinforcing data visibility.

\subsection{User interface and reporting}

\begin{figure*}
  \centering
  \includegraphics[width=.8\linewidth]{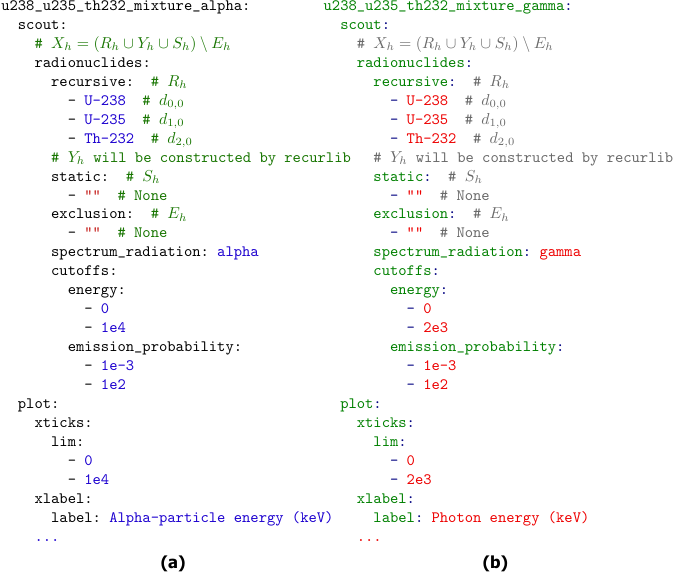}
  \caption{\label{fig:inputSnapshot}%
    Snapshots of a \texttt{RecurLib} user input file configuring the \textbf{a} alpha-particle and \textbf{b} gamma-ray libraries for a mixture of progenitors \textsuperscript{238}U, \textsuperscript{235}U, and \textsuperscript{232}Th.%
  }
\end{figure*}

The current version of \texttt{RecurLib} runs on a command-line interface using the YAML language, offering batch processing capabilities with a human-readable user configuration structure.
At the minimum, specifying the first column vector of a progenitor matrix $\mathbf{R_{h}}$ (\eqRef{eq:manySetsUnionGeneralizedMatrixComputational}), namely
\begin{equation*}
  \begin{bmatrix}
    d_{0,0} \\
    d_{1,0} \\
    \vdots \\
    d_{m,0} \\
  \end{bmatrix}
  \eqPunctComma
\end{equation*}
in the YAML user input file suffices for generating a radionuclide library. For instance, assume that a spectrometry sample is suspected to contain progenitor radionuclides of \textsuperscript{238}U, \textsuperscript{235}U, and \textsuperscript{232}Th. Upon receiving this minimal information on $\mathbf{R_{h}}$ (\figRef{fig:inputSnapshot}), \texttt{RecurLib} constructs a radionuclide subset $\mathbf{X_{h}}$ by computing the progeny matrix $\mathbf{Y_{h}}$ using \eqRef{eq:recurlibFuncGeneral} and the algorithm of energy level feasibility validation (\secRef{sec:nrgLevFeasVal}). Additionally, a static radionuclide matrix $\mathbf{S_{h}}$ and an exclusion radionuclide matrix $\mathbf{E_{h}}$ (\eqRef{eq:rnSubsetMatrix}), if designated at the user prompt, would be merged into $\mathbf{X_{h}}$. The expected radionuclide subset is
\begin{widetext}
\begin{equation*}
  \mathbf{X_{h}} = \begin{bmatrix}
    ^{238}\textnormal{U} & 0 & 0 & 0 & 0 & \cdots & 0 \\
    ^{235}\textnormal{U} & 0 & 0 & 0 & 0 & \cdots & 0 \\
    ^{232}\textnormal{Th} & 0 & 0 & 0 & 0 & \cdots & 0 \\
  \end{bmatrix}
  + \begin{bmatrix}
    0 & ^{234}\textnormal{Th} & ^{234\textnormal{m}}\textnormal{Pa} & ^{234}\textnormal{Pa} & ^{234}\textnormal{U} & \cdots & ^{206}\textnormal{Tl} \\
    0 & ^{231}\textnormal{Th} & ^{231}\textnormal{Pa} & ^{227}\textnormal{Ac} & ^{223}\textnormal{Fr} & \cdots & ^{207}\textnormal{Tl} \\
    0 & ^{228}\textnormal{Ra} & ^{228}\textnormal{Ac} & ^{228}\textnormal{Th} & ^{224}\textnormal{Ra} & \cdots & ^{208}\textnormal{Tl} \\
  \end{bmatrix}
  \eqPunctComma
\end{equation*}
or
\begin{equation*}
  X_{h} = \left\{
  ^{238}\textnormal{U},\,
  ^{234}\textnormal{Th},\,
   ^{234\textnormal{m}}\textnormal{Pa},\,
  \cdots,{ }
  ^{206}\textnormal{Tl},\,
  ^{235}\textnormal{U},\,
  ^{231}\textnormal{Th},\,
   ^{231}\textnormal{Pa},\,
  \cdots,{ }
  ^{207}\textnormal{Tl},\,
  ^{232}\textnormal{Th},\,
  ^{228}\textnormal{Ra},\,
   ^{228}\textnormal{Ac},\,
  \cdots,{ }
  ^{208}\textnormal{Tl}
  \right\}
  \eqPunctDot
\end{equation*}
\end{widetext}

Once constructed, $\mathbf{X_{h}}$ is combined with the corresponding ENSDF nuclear data fetched via the web API of the Live Chart of Nuclide, creating a radionuclide library. The constructed library is then written to external files and visualized in figure formats; the supported file formats are \texttt{.csv}, \texttt{.html}, \texttt{.xml}, \texttt{.tex}, and \texttt{.xlsx} for data reporting and exchange, and \texttt{.pdf}, \texttt{.png}, \texttt{.jpg}, \texttt{.svg}, and \texttt{.emf} for drawing. Generation of each of these formats can be controlled at user input.

\section{Demonstrations}

The library generation performance of \texttt{RecurLib} was investigated using the four actinide decay series, namely the thorium, neptunium, uranium, and actinium series. These families of heavy radionuclides were chosen for their multidisciplinary importance \cite{RN770,RN807}, yet their intricate decay chains and the extensive nuclear data often impede full-fledged spectral analysis. Both alpha-particle and gamma-ray libraries were generated using single progenitors representing each decay chain: \textsuperscript{232}Th, \textsuperscript{237}Np, \textsuperscript{238}U, and \textsuperscript{235}U for the thorium, neptunium, uranium, and actinium series, respectively.

To test the library truncation functionality of \texttt{RecurLib}, the energy ranges for alpha-particle and gamma-ray libraries were set as 0--10,000 keV and 0--2,000 keV, respectively, and the range of emission probabilities as 0.001--100\% for both libraries. The half-life trimming was not applied considering the large spread of half-lives in the actinide decay series, which ranges from $2.94 \times 10^{-7}$ s (\textsuperscript{212}Po)  to $1.40 \times 10^{10}$ y (\textsuperscript{232}Th) except the practically stable nuclide \textsuperscript{209}Bi.

The program execution time, with and without local nuclear data files, was measured for each actinide decay series in quadruplicate and expressed as the mean $\pm$ standard deviation. Additionally, the influence of using a nuclear data registry file (\subfigRef{fig:rnSubsettingAndLineageConst}{a}) on the library generation time was investigated by temporarily disabling the screening feature (\secRef{sec:rnSubsetConst}). A general computational environment was used for the elapsed time measurement: a central processing unit of Intel\textsuperscript{\textregistered} Core\textsuperscript{\texttrademark} i7-11850H Processor, a memory of 32 GB DDR4 2933 MT/s, an operating system of Microsoft Windows 10 Pro, and Internet download speed of 77.7 Mbps.

The effectiveness of the generated radionuclide libraries was tested by performing radionuclide identification using one alpha-particle and eleven gamma-ray spectra encountered in geology and nuclear medicine. I obtained an alpha-particle spectrum of \textsuperscript{225}Ac by measuring a dried 2-$\upmu$L aliquot of \textsuperscript{225}Ac for a live time of 86,400 s (dead time: 0.03\%) using an alpha-particle spectrometer (Model 7401VR Alpha Spectrometer, Canberra Industries Inc., USA). The \textsuperscript{225}Ac activity was approximately 2 kBq which was calculated by gamma-ray spectrometric assay of its immediate daughter \textsuperscript{221}Fr at radioactive equilibrium. The \textsuperscript{225}Ac sample was a product chromatographically purified from its parent \textsuperscript{225}Ra \cite{RN685}, which was produced by proton-induced spallation of \textsuperscript{232}Th at \nobreakdash{TRIUMF} in Canada \cite{RN690}. Two gamma-ray spectra of electron accelerator-produced \textsuperscript{99m}Tc were reproduced from my previous publication \cite{RN515}, both of which were measured using a high-purity germanium (HPGe) detector (GX4018, Canberra Industries Inc., USA) for a live time of 900 s (dead time: 0.2--0.98 \%).

The remaining nine gamma-ray spectra were sourced from an open-source database called Gamma Spectrum DB \cite{RN707}, which is freely available for reuse in line with the permissive GNU General Public License (GPL-3.0). All the nine spectra were measurement results of an HPGe detector with count time of 5--30 min. Of these, six were NORM specimens, two were the therapeutic radionuclides \textsuperscript{225}Ac and \textsuperscript{177}Lu, and one was \textsuperscript{226}Ra, which is in critically short supply for its potential use in \textsuperscript{225}Ac production \cite{RN665}.

\subsection{Actinide decay series}

\begin{figure*}
  \centering
  \includegraphics[width=.98\linewidth]{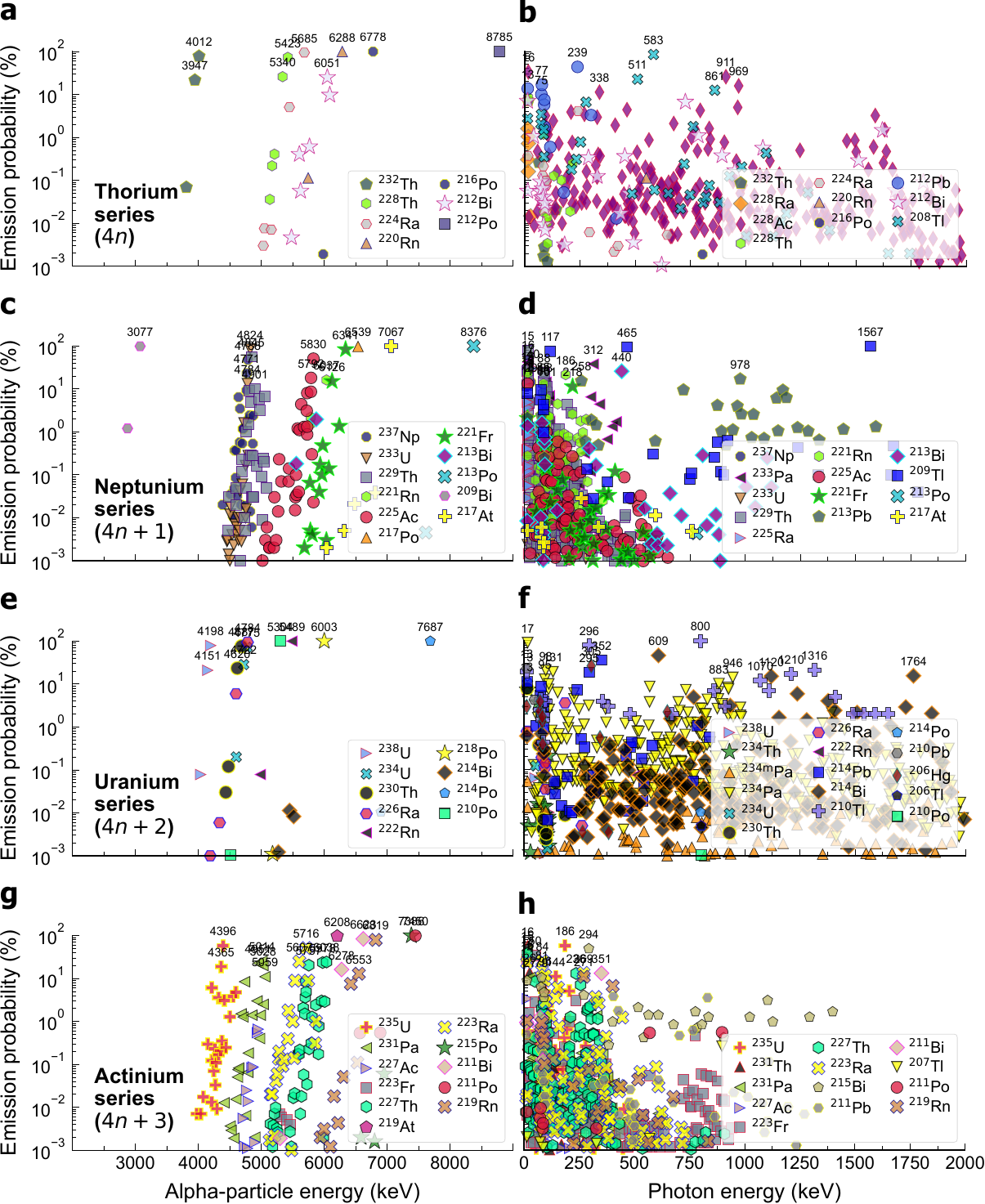}
  \caption{\label{fig:decaySeries}%
    Radionuclide libraries of the four actinide decay series generated by \texttt{RecurLib}. The left and right columns represent alpha-particle and gamma-ray libraries, respectively, for (\textbf{a} and \textbf{b}) thorium, (\textbf{c} and \textbf{d}) neptunium, (\textbf{e} and \textbf{f}) uranium, and (\textbf{g} and \textbf{h}) actinium series. Radiation data points with $\ge$0.001\% emission probabilities are presented, and those with $\ge$10\% are annotated with the corresponding energies.%
  }
\end{figure*}

Radionuclide libraries for the four actinide decay series were accurately generated from their respective single progenitors (\figRef{fig:decaySeries}). The custom-built software \texttt{RecurLib} constructed all the plausible progeny of each actinide decay series as confirmed by lineage diagrams (Supplementary Figures 5--8), and performed the nuclear data retrieval without errors. Its library pruning functionality effectively truncated the alpha-particle and gamma-ray libraries over the designated ranges of energy and emission probability. The library generation time expressed as the mean $\pm$ standard deviation during the first runs were 71.1 $\pm$ 1.0 s, 109.4 $\pm$ 2.6 s, 134.1 $\pm$ 2.8 s, and 115.5 $\pm$ 2.3 s for the thorium ($4n$), neptunium ($4n + 1$), uranium ($4n + 2$), and actinium ($4n + 3$) series, respectively. These were reduced to 6.3 $\pm$ 0.1 s, 15.7 $\pm$ 0.6 s, 19.3 $\pm$ 0.8 s, and 19.5 $\pm$ 0.5 s, respectively, in the second runs where the nuclear datasets archived in the disk at the first runs were reused. This corresponds to execution time decrease of 91.1\%, 85.6\%, 85.6\%, and 83.1\%, respectively. Skipping nonexistent nuclear data using a registry file (\secRef{sec:rnSubsetConst}) were found to reduce the processing time of the second run by 54.9\%, 34.3\%, 24.3\%, and 27.6\%, respectively. \figRef{fig:decaySeries} also demonstrates that the marker registry feature enabled coherent labeling of radionuclides across different library plots.

It should be noted that the emission probabilities shown in \figRef{fig:decaySeries} do not necessarily represent the likelihood of the corresponding radionuclides to be found in a radiation spectrum. This is because a parent radionuclide can breed multiple daughter radionuclides with unique branching fractions to each, and a certain daughter can have a substantially lower branching fraction than its siblings. Moreover, short-lived daughters, regardless of their branching fractions, are typically undetectable unless a time-gated approach \cite{RN747} is applied. In such cases, the amount of measured decay radiation will not be as abundant as its emission probability; ultimately, the propensity of a daughter radionuclide to be included in a spectrum is determined by the combined result of its relative quantity and absolute emission probability.

\subsection{Naturally occurring radioactive materials}

\begin{figure*}
  \centering
  \includegraphics[width=\linewidth]{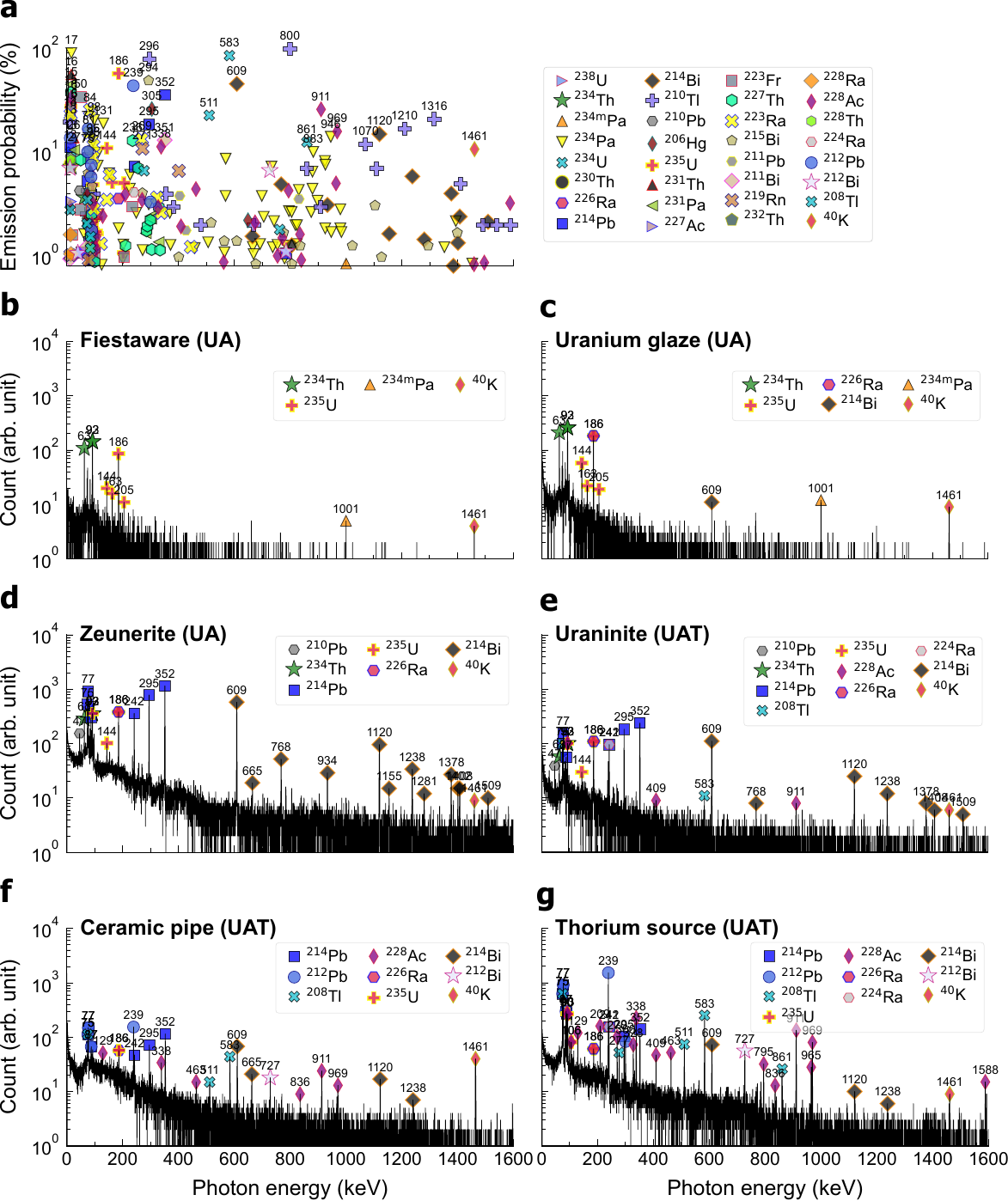}
  \caption{\label{fig:decaySeriesNorm}%
    (\textbf{a}) A \texttt{RecurLib}-generated gamma-ray library and (\textbf{b}--\textbf{g}) its radionuclide identification results. The letters annotated to sample names U, A, and T represent uranium, actinium, and thorium series, respectively.%
  }
\end{figure*}

\FigRef{fig:decaySeriesNorm} shows radionuclides in NORM samples inferred by a \texttt{RecurLib}-generated gamma-ray library. The library (\subfigRef{fig:decaySeriesNorm}{a}) was constructed from input progenitors of \textsuperscript{238}U, \textsuperscript{235}U, \textsuperscript{232}Th, and \textsuperscript{40}K. All samples were associated with \textsuperscript{40}K characterized by its 1460.82-keV gamma radiation (10.66\% emission probability) in addition to the involved actinide decay series. The Fiestaware sample presented in \subfigRef{fig:decaySeriesNorm}{b} is a legacy dinnerware in which triuranium octoxide (U\textsubscript{3}O\textsubscript{8}) of natural origin or in a \textsuperscript{235}U-depleted form was used as a glazing material \cite{RN709}. The presence of uranium series was evidenced by its prominent gamma-emitting progeny \textsuperscript{234}Th (63.29 keV, 3.665\%; 92.38 keV, 2.13\%; and 92.8 keV, 2.1\%) and \textsuperscript{234m}Pa (1001.03 keV, 0.842\%). The actinium series, another major decay chain in uranium-bearing substances, was confirmed by the four gamma-ray peaks of \textsuperscript{235}U (143.765 keV, 10.93\%; 163.357 keV, 5.07\%; 185.713 keV, 57.2\%; and 205.311 keV, 5.03\%). The radionuclic compositions of the Fiestaware and uranium glaze (\subfigRef{fig:decaySeriesNorm}{c}) samples were almost identical, corroborating the use of a uranium glaze in the legacy Fiestaware sample.

The uranium minerals zeunerite and uraninite---the latter commonly known as pitchblende---were found to have similar radionuclic content originating from the uranium and actinium series (\subfigsRef{fig:decaySeriesNorm}{d}{e}), but the uraninite sample was also associated with measurable amounts of the thorium series progeny \textsuperscript{228}Ac, \textsuperscript{224}Ra, and \textsuperscript{208}Tl. The uranium series member \textsuperscript{210}Pb, which can easily migrate from its host because of the relatively long half-life ($t_{1/2}$ = 22.2 y) and the resultant mobility \cite{RN367}, was also detected in these ores. The ceramic pipe and thorium source (\subfigsRef{fig:decaySeriesNorm}{f}{g}) were characterized by the \textsuperscript{232}Th descendants \textsuperscript{212}Pb and \textsuperscript{212}Bi along with the aforementioned ones, suggesting that these samples contain greater amounts of \textsuperscript{232}Th compared with the uraninite specimen.

\subsection{\texorpdfstring{\textsuperscript{225}Ac}{Ac-225}}

\begin{figure*}
  \centering
  \includegraphics[width=\linewidth]{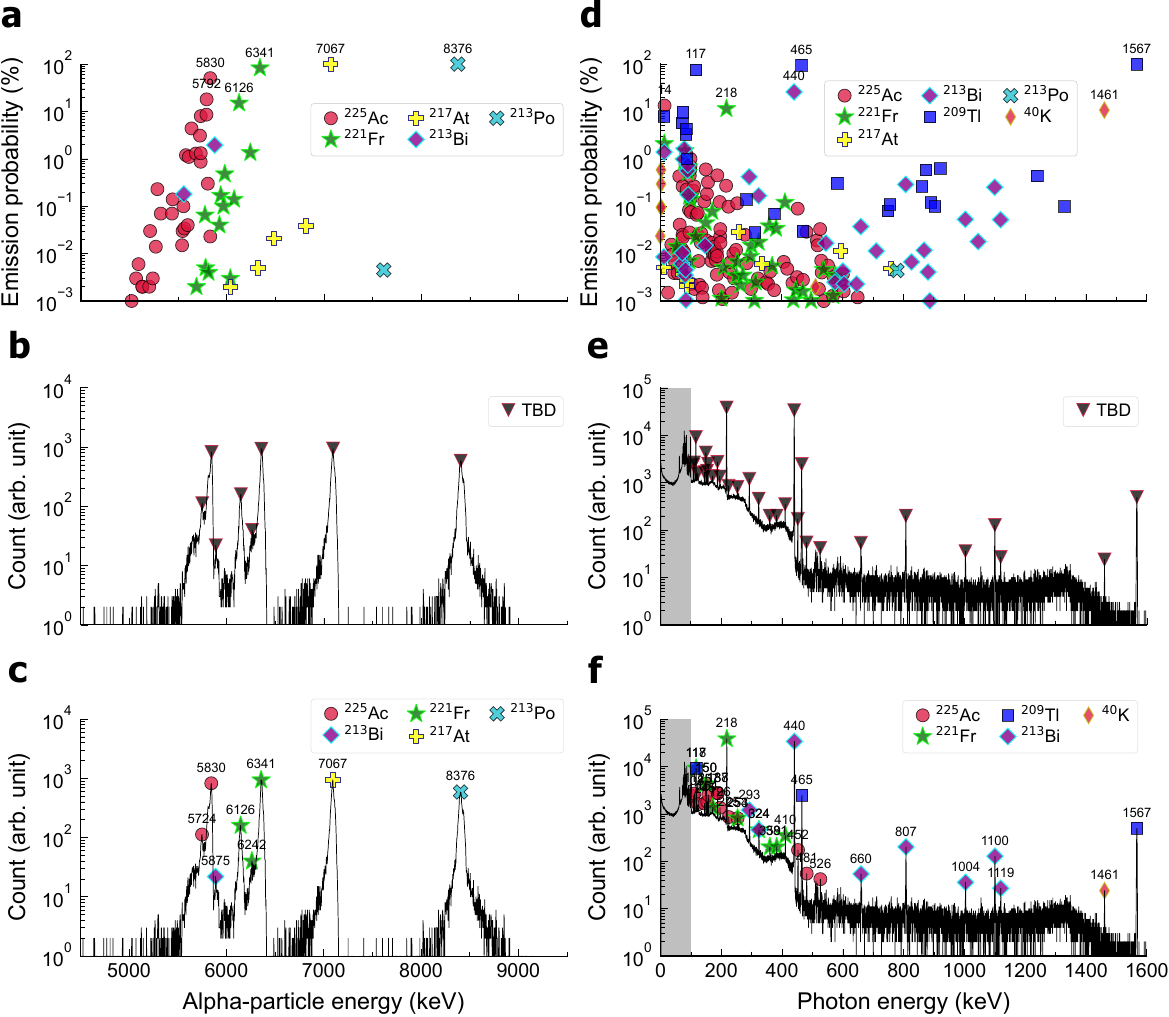}
  \caption{\label{fig:ac225AlphaGamma}%
    (\textbf{a}) Alpha-particle and (\textbf{d}) gamma-ray libraries of \textsuperscript{225}Ac created by \texttt{RecurLib} and (\textbf{b}--\textbf{f}) the corresponding radionuclide identification results. Located spectral peaks whose suspect radionuclides are to be determined are denoted as TBD.%
  }
\end{figure*}

An alpha-particle library of \textsuperscript{225}Ac was formulated using a single progenitor \textsuperscript{225}Ac as \texttt{RecurLib} input (\subfigRef{fig:ac225AlphaGamma}{a}), and a gamma-ray library using progenitors of \textsuperscript{225}Ac and \textsuperscript{40}K (\subfigRef{fig:ac225AlphaGamma}{d}); these libraries were then validated against respective radiation spectra. All the feasible alpha-particle emitters \textsuperscript{225}Ac, \textsuperscript{221}Fr, \textsuperscript{217}At, \textsuperscript{213}Bi, and \textsuperscript{213}Po were identified by the alpha-particle library. The fleet of low-energy gamma rays from \textsuperscript{225}Ac and the prominent gamma rays from \textsuperscript{221}Fr, \textsuperscript{213}Bi, and \textsuperscript{209}Tl were accurately localized by the gamma-ray library.

\subsection{\texorpdfstring{\textsuperscript{226}Ra}{Ra-226} and \texorpdfstring{\textsuperscript{177}Lu}{Lu-177}}

\begin{figure*}
  \centering
  \includegraphics[width=\linewidth]{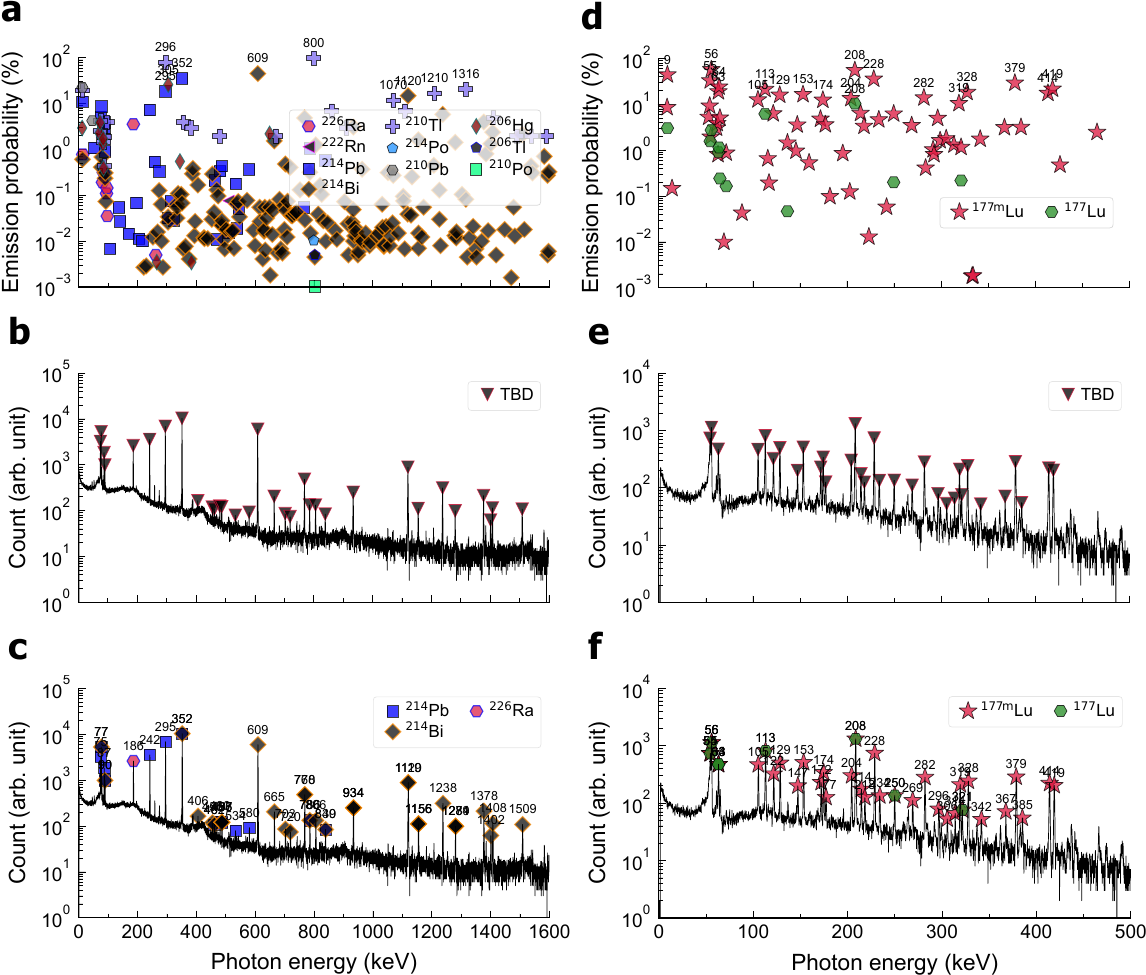}
  \caption{\label{fig:ra226Lu177}%
    Demonstrational radionuclide identification of (\textbf{a}--\textbf{c}) \textsuperscript{226}Ra and (\textbf{d}--\textbf{f}) \textsuperscript{177m,177}Lu samples. The top and bottom rows of each column represent \texttt{RecurLib}-generated gamma-ray libraries and the spotted gamma-ray emitters. The middle row shows gamma-ray spectra where radionuclides responsible for the located peaks are to be determined (TBD).%
  }
\end{figure*}

A \texttt{RecurLib}-created gamma-ray library of \textsuperscript{226}Ra and its radionuclide identification result are presented in Figs. \ref{fig:ra226Lu177}\textbf{a}--\textbf{c}. Radium-226 was the only input progenitor. The presence of \textsuperscript{226}Ra was established from its gamma-ray peak at 186.211 keV (3.565\%) and those from its progeny \textsuperscript{214}Pb and \textsuperscript{214}Bi spreading over the whole displayed energy range. In an unpurified sample, the 186-keV energy region is where \textsuperscript{226}Ra and \textsuperscript{235}U overlap with each other (see Figs. \ref{fig:decaySeriesNorm}\textbf{d}--\textbf{g}). The lack of this spectral interference, coupled with the absence of \textsuperscript{210}Pb and \textsuperscript{234}Th compared with the zeunerite and uraninite data (\subfigsRef{fig:decaySeriesNorm}{d}{e}), suggests that the \textsuperscript{226}Ra species must have been chemically isolated from other nuclides and its progeny grew thereafter. In contrast to Figs. \ref{fig:decaySeriesNorm}\textbf{d}--\textbf{g}, spectral overlaps within \textsuperscript{214}Bi (e.g. 769 keV, 934 keV, and 1120 keV) and between \textsuperscript{214}Pb and \textsuperscript{214}Bi (e.g. 77 keV, 90 keV, 352 keV, 786 keV, and 839 keV) were produced; this was due to the lower emission probability threshold of the \textsuperscript{226}Ra gamma-ray library (0.001\%; see \subfigRef{fig:ra226Lu177}{a}) than that of the NORM gamma-ray library (0.8\%; see \subfigRef{fig:decaySeriesNorm}{a}).

\SubfigRef{fig:ra226Lu177}{d} shows the gamma-ray library of the therapeutic radionuclide \textsuperscript{177}Lu mixed with its radionuclidic impurity \textsuperscript{177m}Lu. A single progenitor of \textsuperscript{177m}Lu was fed to \texttt{RecurLib} with its m4 energy level (970.1757 keV) specified (\secRef{sec:nrgLevFeasVal}), which also included its daughter \textsuperscript{177}Lu in the library. As the longer-lived \textsuperscript{177m}Lu ($t_{1/2}$ = 160.4 d) outnumbers \textsuperscript{177}Lu ($t_{1/2}$ = 6.6 d) with time \cite{RN803}, the considerable presence of \textsuperscript{177m}Lu (\subfigsRef{fig:ra226Lu177}{e}{f}) suggests that the \textsuperscript{177}Lu sample was kept for a certain amount of time after its production.

\subsection{\texorpdfstring{\textsuperscript{99m}Tc}{Tc-99m}}

\begin{figure*}
  \centering
  \includegraphics[width=\linewidth]{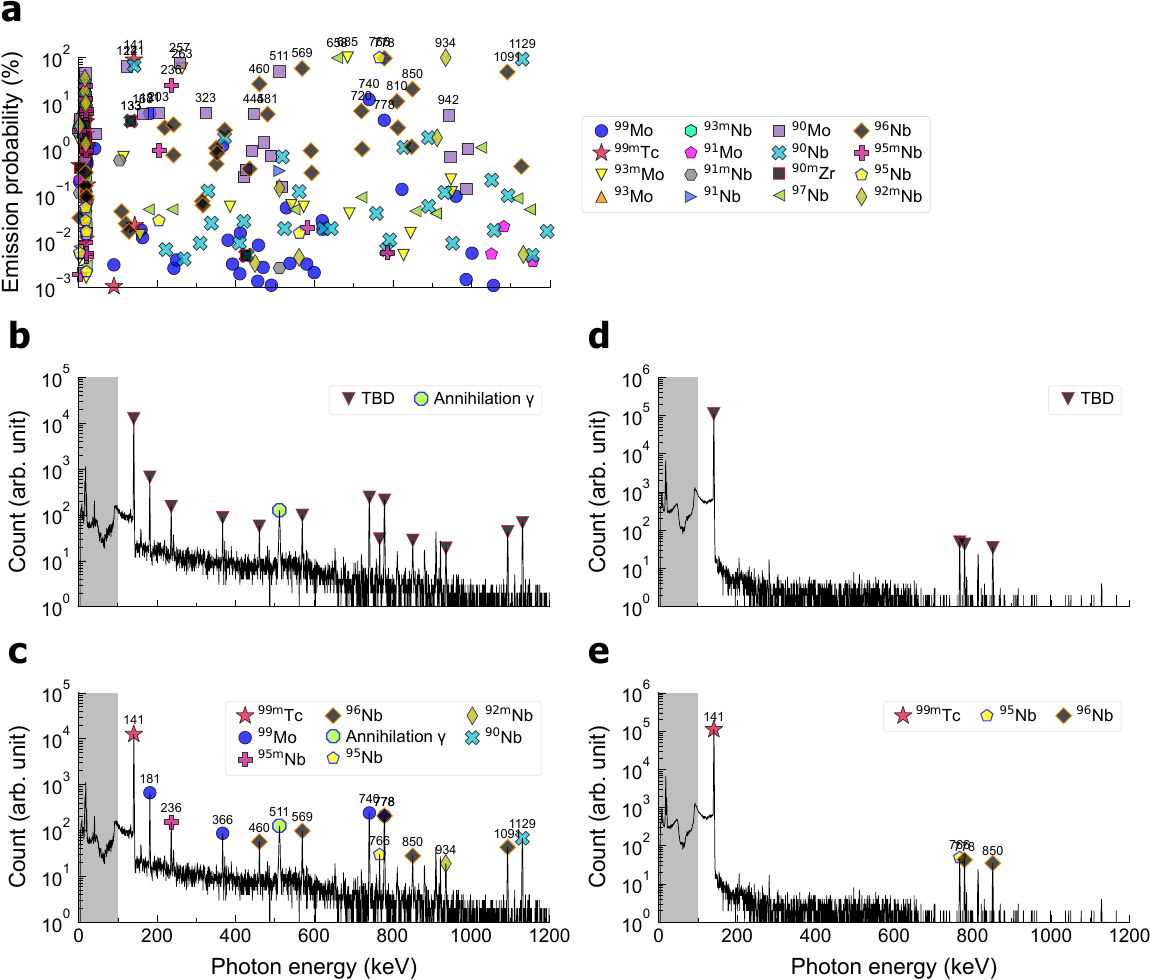}
  \caption{\label{fig:molyTech}%
    (\textbf{a}) A \texttt{RecurLib}-constructed gamma-ray mixture library for a sample containing the products of the photonuclear reaction \textsuperscript{nat}Mo($\upgamma$,$x$). Radionuclide identification results of the samples (\textbf{b} and \textbf{c}) before and (\textbf{d} and \textbf{e}) after \textsuperscript{99m}Tc was chemically separated \cite{RN515}. The acronym TBD stands for ``to be determined''.%
  }
\end{figure*}

\FigRef{fig:molyTech} presents a \texttt{RecurLib}-formulated gamma-ray library of \textsuperscript{99}Mo/\textsuperscript{99m}Tc and its radionuclide identification results. A mixture of six progenitors (\textsuperscript{99}Mo and \textsuperscript{96,95m,95,92m,90}Nb) was assigned as candidate radionuclides \cite{RN366}. Technetium-99m was automatically included in the library as the daughter of \textsuperscript{99}Mo, demonstrating the nuclear isomer inference functionality of \texttt{RecurLib} (\secRef{sec:IsomerInference}). Compared with its original publication \cite{RN515}, \textsuperscript{92m}Nb (934.44 keV, 99.15\%) was additionally spotted. This Nb isomer could be the product of the \textsuperscript{94}Mo($\upgamma$,np)\textsuperscript{92m}Nb reaction \cite{RN366,RN657}.

\section{Discussion}

The success of radionuclide identification in alpha-particle and gamma-ray spectrometry depends significantly on the radionuclide libraries used. Conventional radionuclide library generation involves manually listing candidate radionuclides and collecting their nuclear data, both of which are laborious and susceptible to errors. Through development of a recursive algorithm and its open-source software \texttt{RecurLib}, this research showed that tailored radionuclide libraries containing the latest nuclear data can be created in an automated fashion simply by specifying progenitor radionuclides.

The study presents a conceptual framework for computationally identifying progeny radionuclides and generating customized radionuclide libraries. This approach can streamline the process of radionuclide identification and enhance the accuracy of library data. Input of progenitor radionuclides to the \texttt{RecurLib} software calculates all the plausible progeny, a task typically performed manually by spectrometrists. Additionally, the corresponding nuclear data are automatically fetched from the ENSDF database via the web API of the Live Chart of Nuclides, eliminating the need for manual data retrieval and inspection. This computerized retrieval of nuclear data can facilitate maintaining the accuracy of radionuclide libraries, particularly pertaining to decay radiation and half-life  \cite{RN367,RN776}, for which re-evaluation is frequently carried out. The resulting libraries are essentially custom-built and devoid of unnecessary radionuclides; this tailored nature can reduce the risk of spectral interferences \cite{RN367}.

The task reduction offered by \texttt{RecurLib} can be most clearly shown by the demonstrations of alpha-particle and gamma-ray library construction for the four actinide decay series, namely the thorium, neptunium, uranium, and actinium series. Individual actinide series is associated with 6--8 alpha and 4--6 beta minus decay modes, and the mean numbers of data points are 82.25 for alpha particles and 624 for gamma rays even after a moderate degree of library truncation by energy (0--10,000 keV and 0--2,000 keV for alpha-particle and gamma-ray libraries, respectively) and emission probability (0.001--100\% for both libraries) is applied. These figures will quadruple if all four actinide decay chains are to be encompassed, underscoring the necessity for computerization. Through implementing computerized library generation, such a complex actinide library can now be created within minutes using a single progenitor.

Dual applicability to alpha-particle and gamma-ray spectrometry is one of the defining features of the recursive library construction algorithm. Within the chain of an actinide decay series, the member radionuclides can emit either alpha, beta, or gamma radiation, or a combination of these. A consequence of this is the need to categorize radionuclides into subgroups based on the type of decay radiation within the same decay chain, followed by associating the subgroups with radiation-specific nuclear datasets. Conventional alpha-particle and gamma-ray libraries are therefore developed independently, each necessitating its own master nuclear library. By contrast, the library formulation approach of this study enables constructing both alpha-particle and gamma-ray libraries within the same computational framework using the same nuclear database; this duality was clearly illustrated by the construction of actinide series libraries and the identification of alpha-particle and gamma-ray emitters in \textsuperscript{225}Ac samples. Beta-particle libraries can also be generated by assigning the beta minus particle to the spectrum radiation argument in a \texttt{RecurLib} input file (Supplementary Figures 9--16).

The computational library generation, as designed, is applicable to practically any progenitor radionuclides listed in the ENSDF database, in which over 3,400 nuclides are available at present \cite{RN777}. In this paper, we used radiation spectra encountered in geological and medical sciences to test the validity of \texttt{RecurLib}-constructed radionuclide libraries. This encompassed progenitors spanning from the series-forming actinides \textsuperscript{238}U, \textsuperscript{235}U, and \textsuperscript{232}Th to the medical radionuclides \textsuperscript{225}Ac, \textsuperscript{177}Lu, and \textsuperscript{99m}Tc. The wide radionuclide coverage of computational library generation can prove especially valuable in fields where uncataloged radionuclides are regularly encountered, such as astronomy \cite{RN732,RN727,RN736,RN745}, particle and nuclear physics \cite{RN730,RN729,RN737,RN738,RN744,RN739,RN749,RN746}, and nuclear medicine \cite{RN768,RN690,RN685,RN510,RN742,RN665,RN733,RN734}. In such disciplines, utilizing the computerized library generation algorithm can significantly reduce the time and effort involved in radionuclide identification.

The nuclear data retrieval system of \texttt{RecurLib} ensures data consistency while maintaining access to the latest data. As online databases are regularly updated, nuclear data obtained from the Internet can become inconsistent between spectral analysis runs \cite{RN367}. This potential concern can be obviated in \texttt{RecurLib} because it saves fetched nuclear data to a local disk and reuses them from the second occasion onward. Repeated use of the same nuclear data files means data consistency; conversely, deleting existing files will trigger online data retrieval and update the datasets if new revisions are available.

Qualifying minor radiation peaks that are not employed for activity calculation can provide spectral evidence on the radionuclidic compositions, especially when a radionuclide having multiple radiation emissions is present. In our demonstrational radionuclide identification, \textsuperscript{225}Ac entailed fifteen such gamma-ray peaks from itself over the energy range of 108.4--526.1 keV, and eight peaks from its immediate progeny \textsuperscript{221}Fr over 117.8--410.4 keV except the major peak at 218 keV. However, the gamma-ray spectra of \textsuperscript{225}Ac samples in published literature \cite{RN768,RN690,RN665} rarely have annotations for these peripheral peaks; this may be due to the triviality of the minor peaks in terms of \textsuperscript{225}Ac activity quantification, but largely because of their extensive amount that requires laborious manual registration by users. Another example is the \textsuperscript{177}Lu sample, which exhibited 38 detectable gamma-ray peaks of \textsuperscript{177m}Lu over the range of 55.15--418.54 keV. Unless qualified, some of these peaks may be mistakenly attributed to false radionuclides; if properly ascribed, they would serve to corroborate the presence of the actual radionuclides. Through the demonstrations of \textsuperscript{225}Ac and \textsuperscript{177}Lu samples, this paper showed that minor peaks can easily be qualified by utilizing the automated nuclear data retrieval feature of \texttt{RecurLib}. This also highlights the applicability of \texttt{RecurLib} in radionuclidic composition assessment.

Recent years have seen increasing utilization of pattern recognition technologies in radionuclide identification, particularly for radiation portal monitors (RPMs) where rapid detection is prioritized. Paff and colleagues \cite{RN779} applied a spectral angle mapper (SAM) method for in situ detection of radionuclides at RPMs. The SAM algorithm is used to determine the presence of suspected radionuclides by comparing the spectral similarity between test and reference spectra. While this approach provides a means of fast radionuclide identification, radionuclides that are not commonly encountered at RPMs and thus having no reference data would not be recognized. This contrasts with the conventional library-guided method, which does not necessitate comparison spectra.

Several research groups \cite{RN780,RN781,RN782} devised radionuclide identification algorithms for RPMs employing artificial neural networks (ANNs). These studies addressed the initial spectral training efforts and reported promising performance with reduced identification time. Considering the high priority placed on speed in radionuclide identification at RPMs, ANN-assisted pattern recognition algorithms present a compelling opportunity to supplant the conventional library-guided approach. However, a considerable amount of learning seems to be necessary until a satisfactory level of generalization capability is acquired, given the large numbers of known radionuclides and their decay radiation. In contrast, laboratory spectral assays are focused more on dealing with radiation spectra lacking corresponding templates and those featuring a number of multiplet peaks that can be resolved by spectral deconvolution, for which library-guided radionuclide identification methods remain an essential tool. When ANNs with comprehensive generalization capabilities become available, hybridizing the library- and ANN-driven approaches could potentially enhance the overall performance of radionuclide identification in both RPMs and laboratory settings.

A software functionality allowing users to select which nuclear database to use can expand the applicability of \texttt{RecurLib}. Currently, ENSDF is the only nuclear database accessible within \texttt{RecurLib}. While ENSDF is arguably the gold standard source of nuclear datasets, access to other databases can provide a greater degree of customization in radionuclide libraries.

Another feature that needs to be added is a graphical user interface (GUI). At present, \texttt{RecurLib} runs only on a command-line interface (CLI) based on the YAML language. Although the CLI offers straightforward ways of batch processing and software integration, some users may find CLI environments unfamiliar or uneasy to use; to accommodate these needs, GUI development is under consideration for future updates.

\section{Conclusion}

This study developed a theoretical framework and its open-source software \texttt{RecurLib} for fully automated generation of purpose-built radionuclide libraries with enhanced data reliability. The demonstrational alpha-particle and gamma-ray libraries were generated within minutes and effectively identified radionuclides in twelve multidisciplinary spectrometry samples, illustrating the efficiency and practicality of the proposed algorithm. This computational approach can simplify the process of library-guided radionuclide identification and facilitate accommodating virtually any radionuclides available in the ENSDF database, expanding the applicability of radiation spectrometry across diverse disciplines.

\section*{CRediT authorship contribution statement}

\CRediTAuth{\authI:} Conceptualization, Methodology, Software, Validation, Formal analysis, Investigation, Resources, Data Curation, Writing - Original Draft, Writing - Review \& Editing, Visualization, Project administration, Funding acquisition

\section*{Declaration of competing interest}

The author declares no competing interests.

\section*{Data availability}

The alpha-particle spectrum of \textsuperscript{225}Ac is available from the author on reasonable request. All gamma-ray spectra presented in this paper can be accessed online, either copyrighted \cite{RN515} or within the public domain \cite{RN707}.

\section*{Code availability}

The codebase of \texttt{RecurLib} is openly accessible and maintained in a GitHub repository \cite{RN809} and its snapshot corresponding to the work of this paper has been archived in Zenodo \cite{RN810}. All these resources are available under the permissive MIT license.

\section*{Acknowledgments}

This research was supported by JSPS KAKENHI Grant Numbers JP22K20880 and JP23K17147. The author prepared this paper during his research visit to TRIUMF in Canada, which was supported by the University of Tokyo through its UTokyo Global Activity Support Program for Young Researchers. TRIUMF receives federal funding via a contribution agreement with the National Research Council of Canada.

The author expresses gratitude to Youichiro Wada (University of Tokyo), Takashi Nakano (Osaka University), and Paul Schaffer (TRIUMF/University of British Columbia/Simon Fraser University/Osaka University) for their provision of \textsuperscript{225}Ac, and to Shogo Higaki (University of Tokyo) for his support in conducting \textsuperscript{225}Ac alpha-particle spectrometry. Special appreciation is extended to Matthias Rosezky for the public release of his curated gamma-ray spectrum database Gamma Spectrum DB, based on which most of the software demonstrations in this paper were carried out. The Live Chart of Nuclides and the ENSDF database were instrumental in developing the \texttt{RecurLib} software.

\bibliography{ms}

\end{document}


{\noindent\large\textbf{Supporting Information}}

\title{\theTitle}
\author{\authI}
\email[]{\authIEmail}
\affiliation{\addrUTokyoISC}
\maketitle

\begin{figure}[h]
  \centering
  \includegraphics[width=\linewidth]{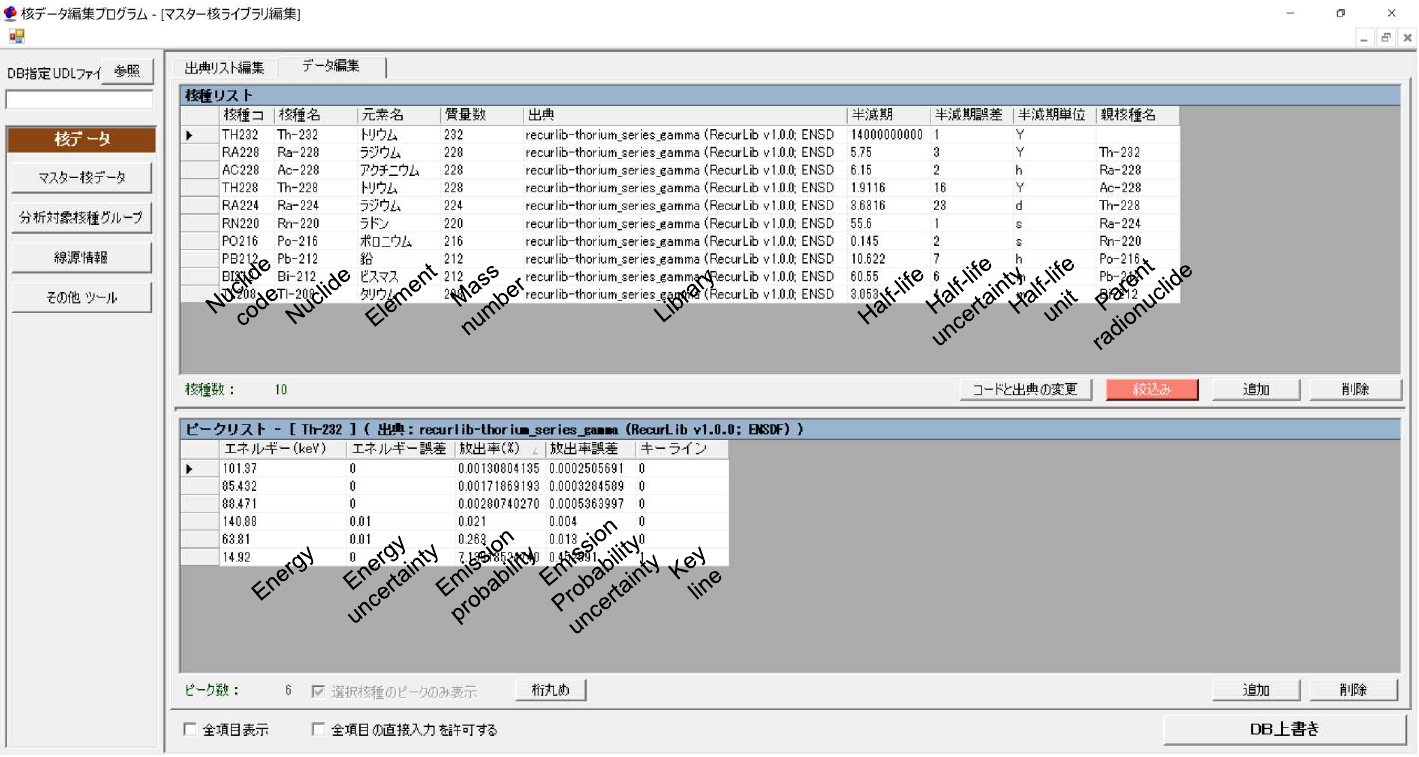}
  \caption{\label{fig:gexpThGamma}%
    A \texttt{RecurLib}-generated gamma-ray library of thorium series imported into \texttt{Nuclear Data Editor}, \texttt{Gamma Explorer} v1.74 (Mirion Technologies (Canberra) KK, Japan).%
  }
\end{figure}

\begin{figure}
  \centering
  \includegraphics[width=\linewidth]{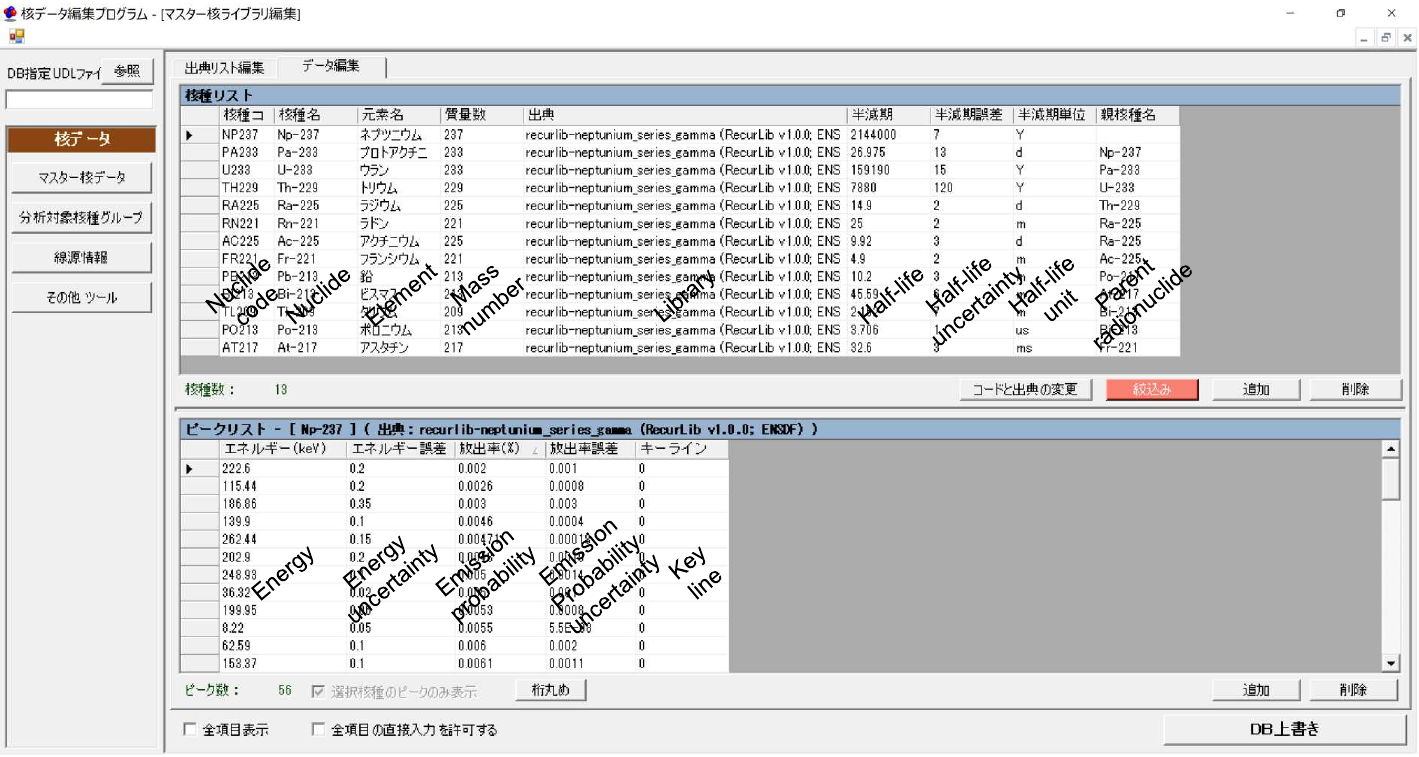}
  \caption{\label{fig:gexpNpGamma}%
    A \texttt{RecurLib}-generated gamma-ray library of neptunium series imported into \texttt{Nuclear Data Editor}, \texttt{Gamma Explorer} v1.74 (Mirion Technologies (Canberra) KK, Japan).%
  }
\end{figure}

\begin{figure}
  \centering
  \includegraphics[width=\linewidth]{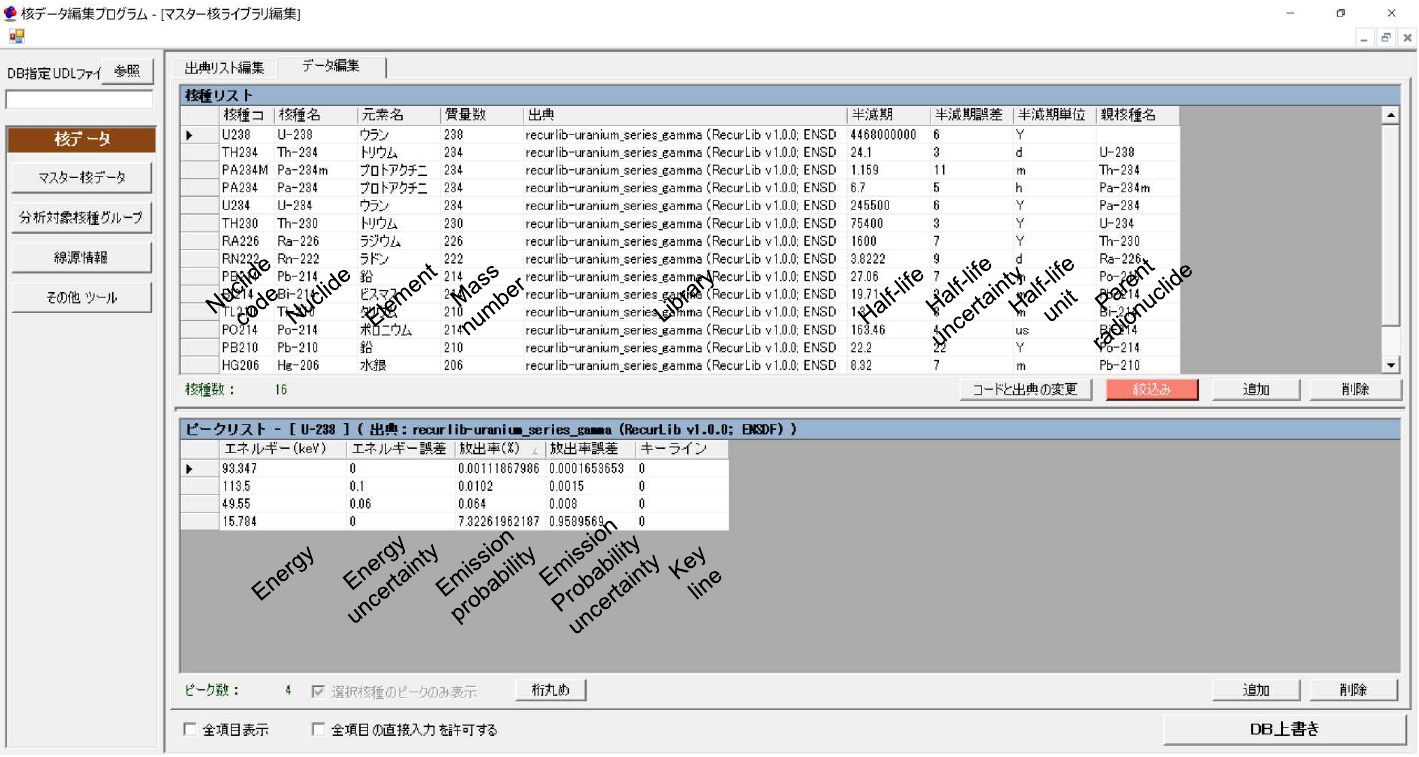}
  \caption{\label{fig:gexpUGamma}%
    A \texttt{RecurLib}-generated gamma-ray library of uranium series imported into \texttt{Nuclear Data Editor}, \texttt{Gamma Explorer} v1.74 (Mirion Technologies (Canberra) KK, Japan).%
  }
\end{figure}

\begin{figure}
  \centering
  \includegraphics[width=\linewidth]{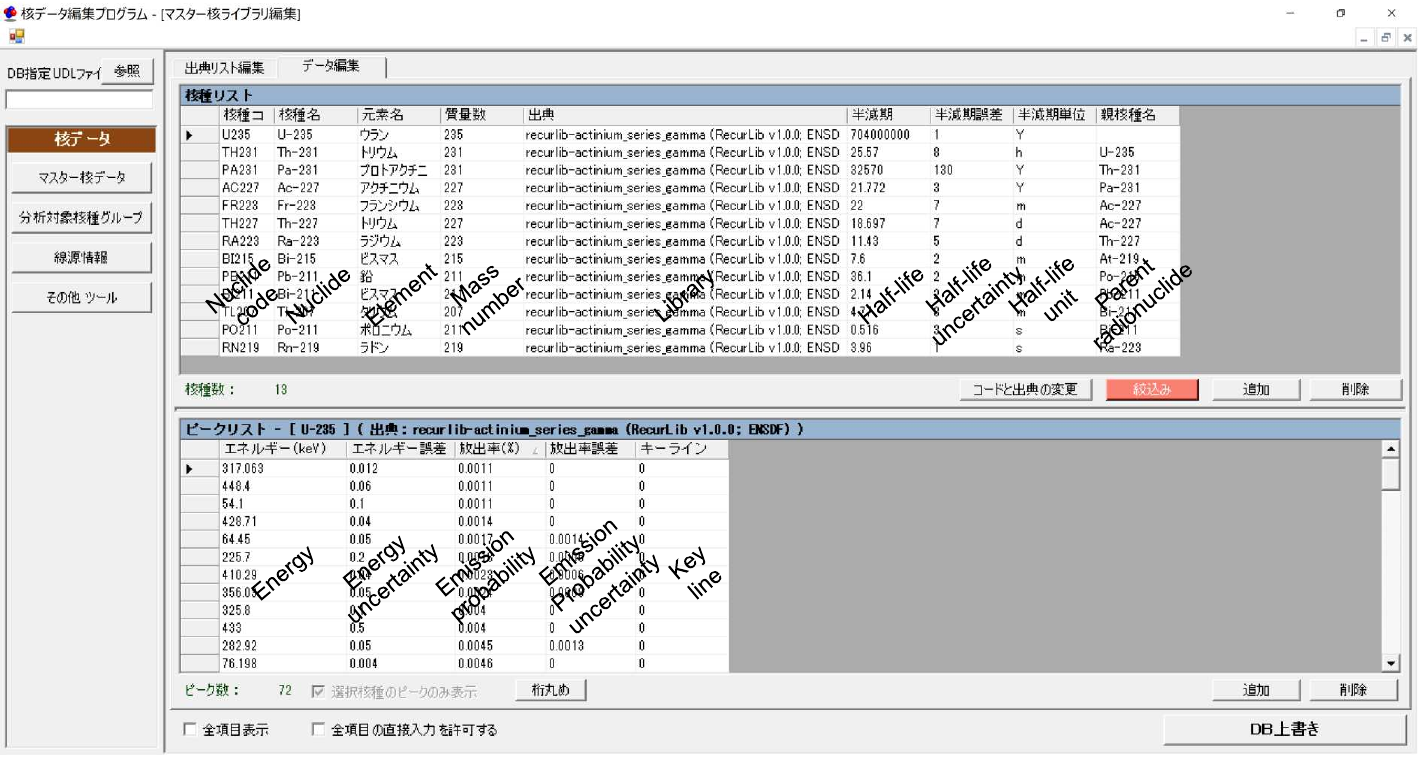}
  \caption{\label{fig:gexpAcGamma}%
    A \texttt{RecurLib}-generated gamma-ray library of actinium series imported into \texttt{Nuclear Data Editor}, \texttt{Gamma Explorer} v1.74 (Mirion Technologies (Canberra) KK, Japan).%
  }
\end{figure}

\begin{figure}
  \centering
  \includegraphics[width=0.44\linewidth]{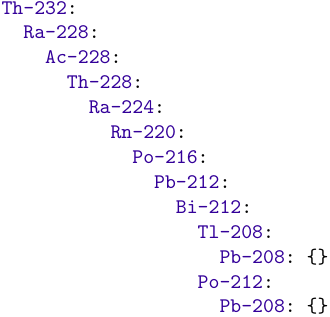}
  \caption{\label{fig:linTh}%
    A \texttt{RecurLib}-generated lineage diagram of thorium series.%
  }
\end{figure}

\begin{figure}
  \centering
  \includegraphics[width=0.5\linewidth]{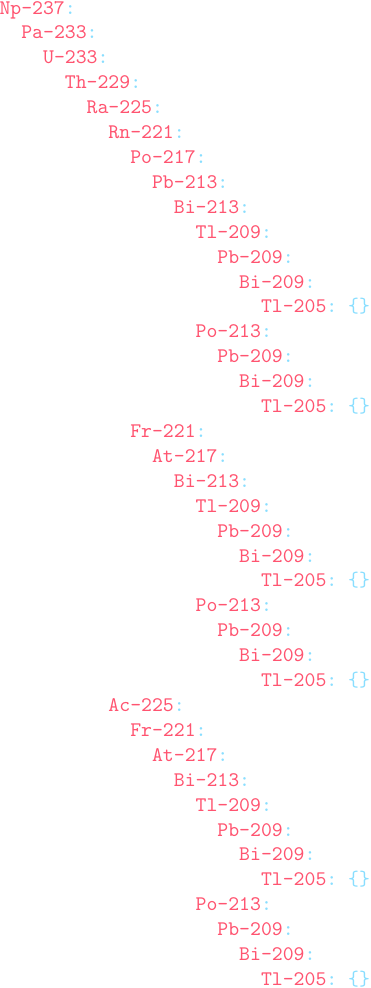}
  \caption{\label{fig:linNp}%
    A \texttt{RecurLib}-generated lineage diagram of neptunium series.%
  }
\end{figure}

\begin{figure}
  \centering
  \includegraphics[width=0.57\linewidth]{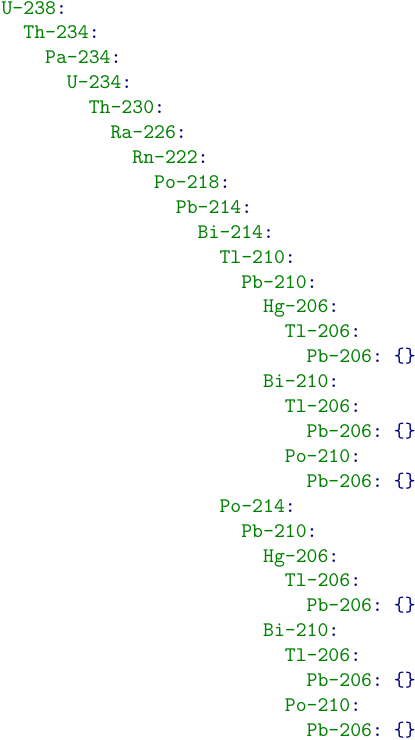}
  \caption{\label{fig:linU}%
    A \texttt{RecurLib}-generated lineage diagram of uranium series.%
  }
\end{figure}

\begin{figure}
  \centering
  \includegraphics[width=0.48\linewidth]{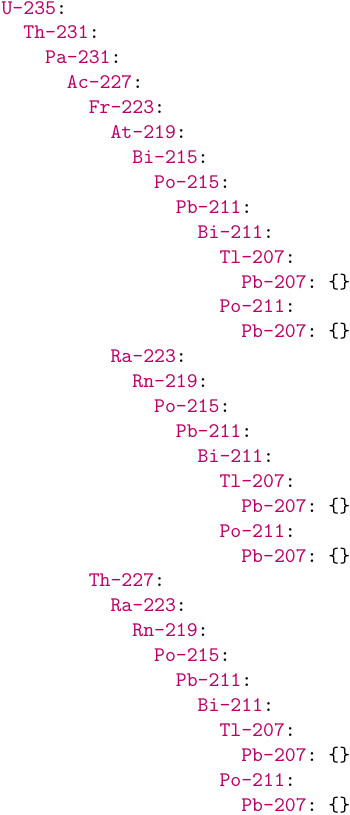}
  \caption{\label{fig:linAc}%
    A \texttt{RecurLib}-generated lineage diagram of actinium series.%
  }
\end{figure}

\begin{figure}
  \centering
  \includegraphics[width=0.8\linewidth]{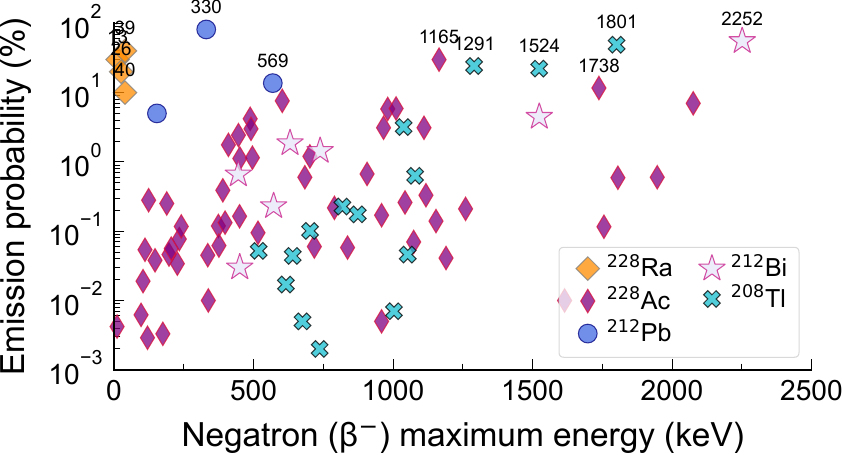}
  \caption{\label{fig:betaLibTh}%
    A \texttt{RecurLib}-generated beta-particle library of thorium series.%
  }
\end{figure}

\begin{figure}
  \centering
  \includegraphics[width=0.8\linewidth]{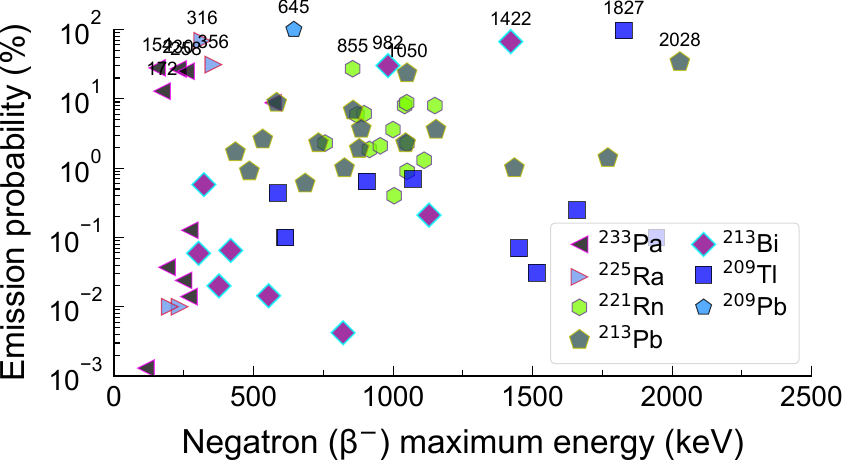}
  \caption{\label{fig:betaLibNp}%
    A \texttt{RecurLib}-generated beta-particle library of neptunium series.%
  }
\end{figure}

\begin{figure}
  \centering
  \includegraphics[width=0.8\linewidth]{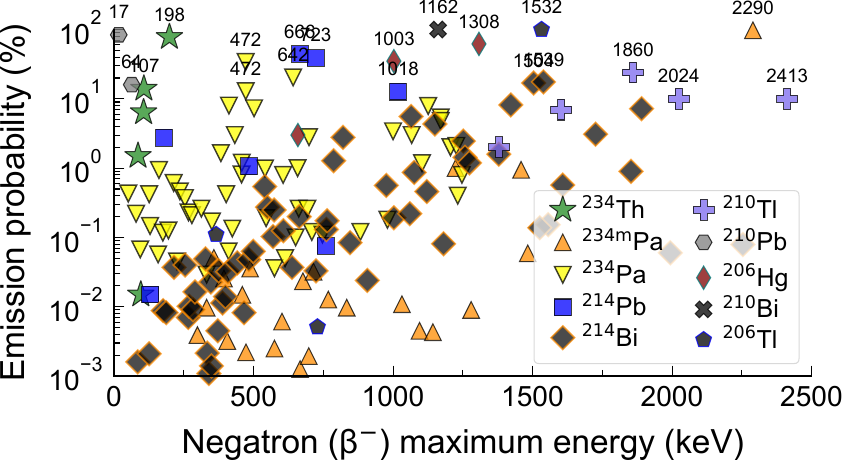}
  \caption{\label{fig:betaLibU}%
    A \texttt{RecurLib}-generated beta-particle library of uranium series.%
  }
\end{figure}

\begin{figure}
  \centering
  \includegraphics[width=0.8\linewidth]{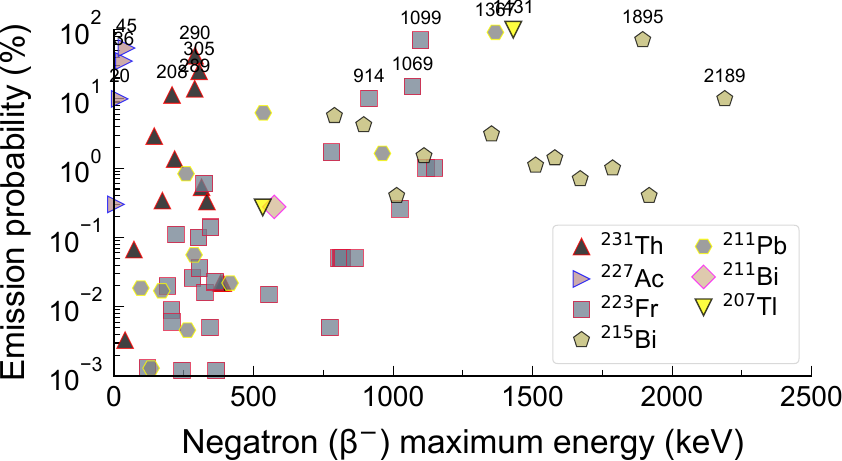}
  \caption{\label{fig:betaLibAc}%
    A \texttt{RecurLib}-generated beta-particle library of actinium series.%
  }
\end{figure}

\begin{figure}
  \centering
  \includegraphics[width=0.8\linewidth]{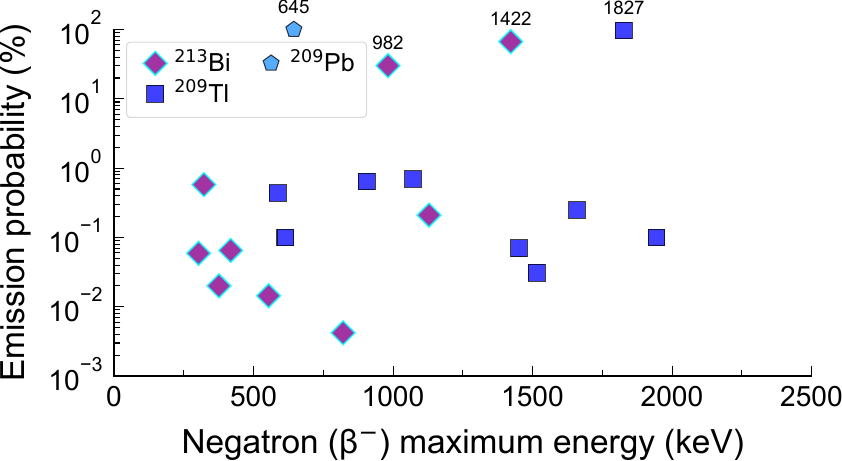}
  \caption{\label{fig:betaLibAc225}%
    A \texttt{RecurLib}-generated beta-particle library of \textsuperscript{225}Ac.%
  }
\end{figure}

\begin{figure}
  \centering
  \includegraphics[width=0.8\linewidth]{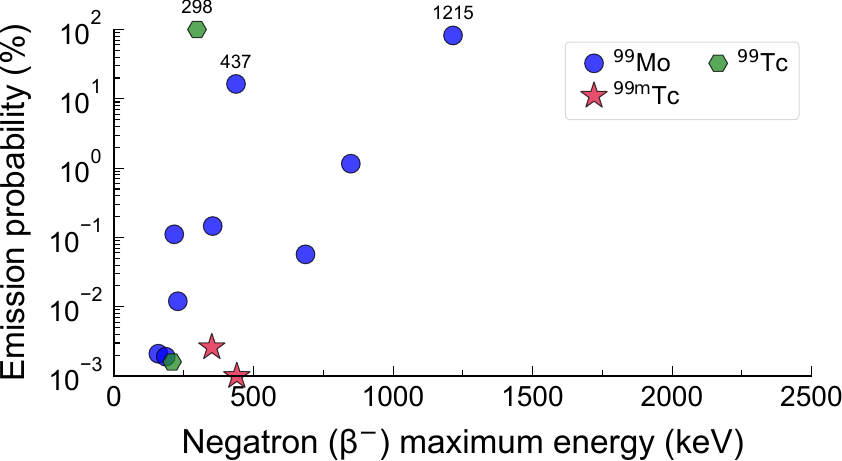}
  \caption{\label{fig:betaLibMolyTech}%
    A \texttt{RecurLib}-generated beta-particle library of \textsuperscript{99}Mo.%
  }
\end{figure}

\begin{figure}
  \centering
  \includegraphics[width=0.8\linewidth]{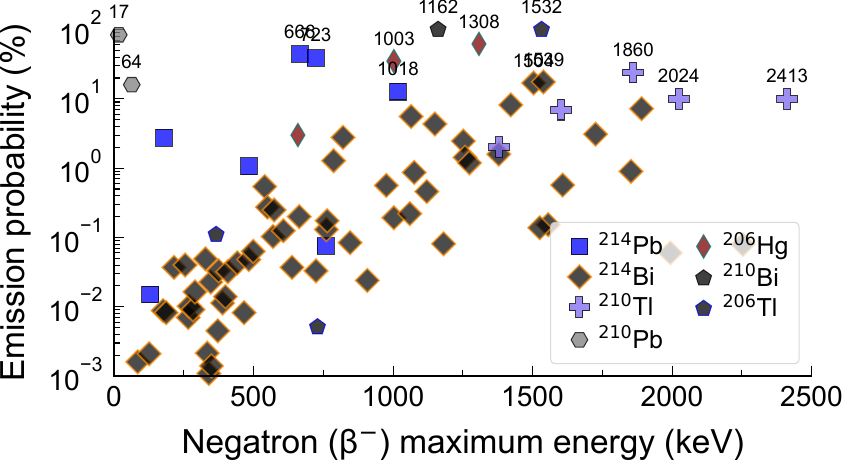}
  \caption{\label{fig:betaLibRa}%
    A \texttt{RecurLib}-generated beta-particle library of \textsuperscript{226}Ra.%
  }
\end{figure}

\begin{figure}
  \centering
  \includegraphics[width=0.8\linewidth]{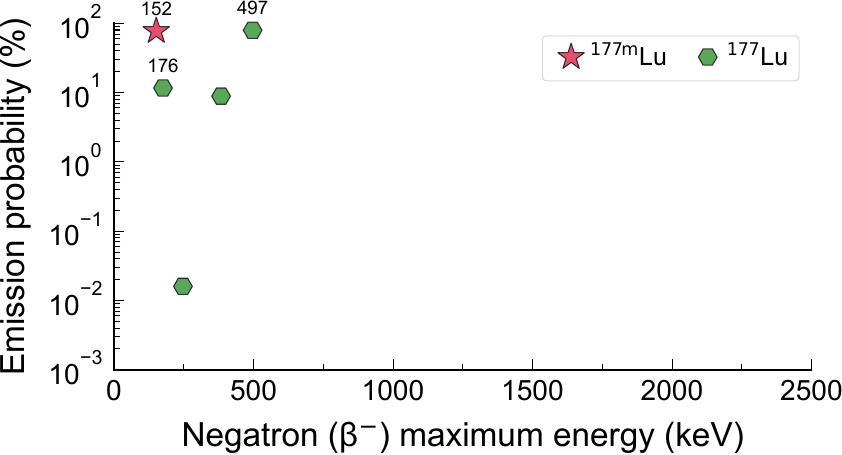}
  \caption{\label{fig:betaLibLu177}%
    A \texttt{RecurLib}-generated beta-particle library of \textsuperscript{177m}Lu.%
  }
\end{figure}